\documentclass[doublespacing]{elsart}

\usepackage{icarus}

\usepackage{natbib}

\bibpunct{(}{)}{;}{a}{,}{,}

\usepackage{epsfig}
\usepackage{graphicx}

\usepackage{color}%

\def\cora#1{\textbf{\textcolor{black}{\rm#1}}}   
\def\corb#1{\textbf{\textcolor{black}{\rm#1}}}   

\def\cord#1{\textbf{\textcolor{black}{\rm#1}}}
\def\core#1{\textbf{\textcolor{black}{\rm#1}}}

\def\ie{{ i.e.}}  
\def\eg{{ e.g.}}  
  
\def\deg{$^{\circ}$}
\def\nh3{ $\mbox{NH}_3$}
\def\ch4{$\mbox{CH}_4$}
\def\h2o{ $\mbox{H}_2\mbox{O}$}
\def\c2h2{ $\mbox{C}_2\mbox{H}_2$}

\newcommand{\beginsupplement}{%
        \setcounter{table}{0}
        \renewcommand{\thetable}{S\arabic{table}}%
        \setcounter{figure}{0}
        \renewcommand{\thefigure}{S\arabic{figure}}%
     }

\newcommand{\beginappendix}{%
        \setcounter{table}{0}
        \renewcommand{\thetable}{A\arabic{table}}%
        \setcounter{figure}{0}
        \renewcommand{\thefigure}{A\arabic{figure}}%
     }

\usepackage{lineno}

\begin{document}

\begin{frontmatter}



\title{Saturn's aurora observed by the Cassini camera at visible wavelengths.}


\author[caltech]{Ulyana A. Dyudina}, 
\author[caltech]{Andrew P. Ingersoll},
\author[caltech]{Shawn P. Ewald}, 
\author[arizona]{Danika Wellington}

\address[caltech]{Division of Geological and Planetary Sciences, 
 150-21 California Institute of Technology,  Pasadena, CA 91125 (U.S.A.)}
\address[arizona]{School of Earth and Space Exploration, Arizona State University, ISTB4 Rm 795 781 Terrace Rd
Tempe, AZ 85287-6004 (U.S.A.)}

\end{frontmatter}



\begin{flushleft}
Number of text pages: \pageref{lastpage}, 
Number of tables: 1 
Number of figures: \ref{fig:image_areas_spectra}\\
\end{flushleft}


\begin{pagetwo}{   Saturn's Visible Aurora          }

Ulyana A. Dyudina \\
150-21 Caltech, Pasadena, CA 91125, USA \\
\\
E-mail: ulyana@gps.caltech.edu\\
Phone: (626)395-6824 \\
Fax: (626)585-1917

\end{pagetwo}

\begin{abstract}

\cora{The first observations of Saturn's visible-wavelength aurora were made by the Cassini camera.
The aurora was observed between 2006 and 2013 in the northern and southern hemispheres.}
The color of the aurora changes from pink at a few hundred km above the horizon to purple at 1000-1500 km above the horizon.
The spectrum observed in 9 filters spanning wavelengths from 250 nm to 1000 nm has a prominent H-alpha line and roughly agrees with laboratory simulated auroras. 
Auroras in both hemispheres vary dramatically with longitude.
Auroras form bright arcs 
\core{between 70{\deg} and 80{\deg} latitude north and between 65{\deg} and 80{\deg} latitude south}
\cora{, which sometimes  spiral around the pole, and sometimes form} double arcs. 
A large 10,000-km-scale longitudinal brightness structure  persists for more than 100 hours.
This structure rotates approximately together with Saturn.
On top of the large steady structure, the auroras brighten suddenly on the timescales of a few minutes.
These brightenings repeat with a period of $\sim$1 hour.
Smaller, 1000-km-scale structures may move faster or lag behind Saturn's rotation on timescales of tens of minutes.
\cord{The persistence of nearly-corotating large bright longitudinal structure in the auroral oval seen in two movies spanning 8 and 11 rotations gives an estimate on the period of 10.65 $\pm$ 0.15 h for 2009 in the northern oval and 10.8$\pm$ 0.1 h for 2012 in the southern oval.}
The 2009 north aurora period is close to the north branch of Saturn Kilometric Radiation (SKR) detected at that time.


\end{abstract}



\section{Introduction.}
\label{sec:introduction}

Before Cassini arrived at Saturn, Saturnian aurora was observed \cora{in both} UV and infrared (IR) wavelengths\cora{, as reviewed by \cite{saturn10b_aurora}}.  
Substantial auroral research \cora{at Saturn} has been \cora{undertaken since} then.
This includes Cassini UV and IR movies \cora{\citep{pryor11,carbary12, badman11}}, radio data \citep{provan13, cowley13}, and magnetospheric particle maps \citep{lamy13}.
Also Earth-based observations of UV aurora and solar wind were performed \cora{\citep{clarke09,grodent10}}.
Good summaries of the recent discoveries are given by \cite{lamy13} \cora{and \cite{grodent14}}.

\cord{Particles precipitating to the upper atmosphere of Saturn form circumpolar auroral ovals, as on Earth and Jupiter, however the origin of these particles is controversial.}
\cora{Correlation of auroral dynamics with the solar wind suggests that the aurora is at least partly driven by solar wind, \core{possibly with the boundary between open and closed magnetic field lines projecting to the main auroral oval,} as on Earth.
However, solar-wind-independent variability of aurora 
suggests that the aurora is also driven by internal disturbances of the ion-loaded magnetosphere \citep[\eg, modeled by][]{cowley04, goldreich07,gurnett07}, as on Jupiter.
This mass loading is produced by volcanic activity on Jupiter's and Saturn's moons. 
UV and infrared aurora and SKR show some correlation with solar wind \citep{ gerard05, stallard12, nichols14}, however the aurora also varies independently \citep{clarke09}.
This leaves the question of the origin of Saturn's aurora open for observers and modelers.}

\cora{This paper is the first to report detection of aurora in visible light by the Cassini camera in 2006. 
Visible-light images and movies by the Cassini camera show aurora at unprecedented spatial resolution as fine as tens of km per pixel, and also at unprecedented time resolution as fine as one minute.
\core{Previous detections of auroral ovals and 500-km-scale arcs and spots in UV had spatial resolution of hundreds of km \citep{gerard04,stallard08, grodent11, radioti14}.}
Visible aurora is harder to detect than UV and IR aurora because in visible wavelengths daylight interferes with the auroral light.
The auroral brightness is only 10$^{-6}$--10$^{-5}$ \core{(10$^{-4}$--10$^{-3}$ per cent)} of the brightness of sunlit dayside of Saturn.
Because of that, visible aurora can only be observed on} the night side. 
Only spacecraft near Saturn can observe the night side and detect visible aurora.
\cora{Similar restrictions apply to Jupiter, where \core{visible} aurora was first observed in Galileo spacecraft images \citep{vasavada99}}

Here we present all the visible auroral observations starting with Cassini's arrival at Saturn in 2004 until March 2014.
Section \ref{sec:geometry} describes Saturn's aurora \cora{morphology and location}.
Section \ref{sec:timing} reveals the timescales of auroral variations. 
Section \ref{sec:spectrum} reports the spectrum of the auroras obtained with different filters on the camera, together with the vertical structure of aurora.
Details of image processing and a list of all auroral detections are presented in the Appendix.

\section{Auroral Morphology and Location}
\label{sec:geometry}

\begin{figure}[htbp]
 \resizebox{9in}{!}{\includegraphics{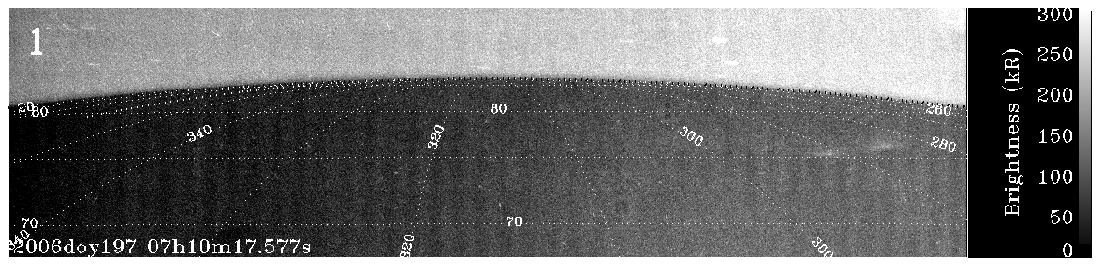}} 
    \vspace{.1in}
    \resizebox{9in}{!}{\includegraphics{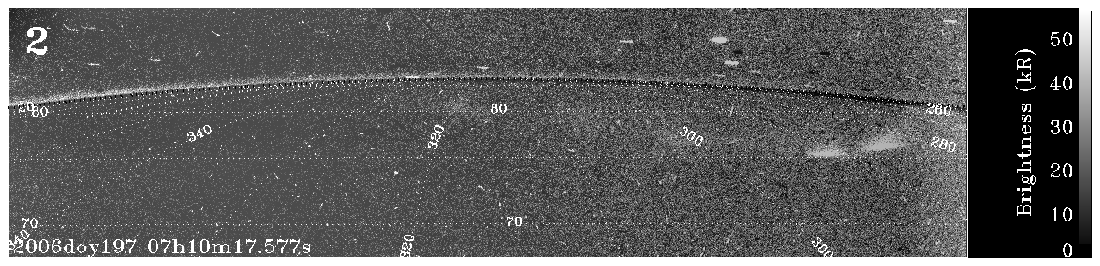}} 
    \resizebox{9in}{!}{\includegraphics{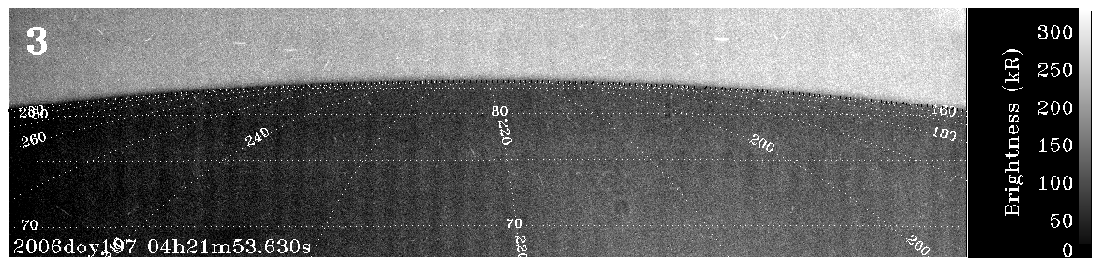}} 
    \resizebox{9in}{!}{\includegraphics{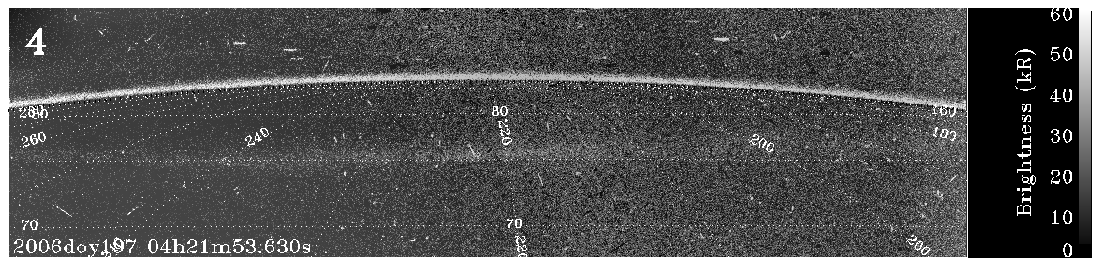}}
    \caption{
Images of Saturn aurora on July 16 (which is Day of Year, or DOY 197), 2006.
Panels 1 and 3 show original images converted to brightness units of Rayleighs ($1{R} =10^{10}(photons)m^{-2}s^{-1}$, see the kilo-Rayleigh (kR) scale bar on the right).
A constant brightness value was subtracted from each image to account for stray light in the camera. 
The value was chosen to maximize the image contrast.
Panels 2 and 4 show the same images with the average of three other similar images ({\ie} "background image") subtracted to reveal aurora changing from image to image.
A west longitude and planetocentic System III latitude grid overlays the images.
The date and time are labeled in the lower left of each panel. 
Each image was taken using a broadband filter spanning the entire range of visible wavelengths  250 to 1000 nm, which is the CL1+CL2 filter combination \citep[see filter details in][]{porco04}. 
The filter shape is shown in Fig. \ref{fig:image_areas_spectra}. 
} 
    \label{fig: unprojected} 
\end{figure}

Figure \ref{fig: unprojected} shows the discovery images of Saturn's visible aurora taken on July 16, 2006.
\renewcommand{\baselinestretch}{1.}
\renewcommand{\baselinestretch}{2.}
The Cassini camera is observing the north polar area at night. 
In the raw images (Panels 1 and 3), Saturn's limb is a dark silhouette against the brighter background of the "clear sky". 
The "clear sky" shows star trails and some opacity source behind the planet, possibly E-ring material, which \cora{will be} the subject of a separate research. 
The raw images (Panels 1 and 3) are processed for noise reduction and shown again in Panels 2 and 4, respectively.

\subsection{Image Processing}

\cora{To enhance the contrast of the faint aurora on the noisy background we used two techniques.
First, we took advantage of the variable nature of the aurora during the multiple-frame movies.
For each movie, we constructed the average image from all movie frames not containing obvious aurora, \ie,  the background image. 
Then we subtracted that background image from each frame in the movie.
This left only the variable part of the brightness, which is predominantly aurora.
This technique removes stray light in the camera, detector defects such as vertical stripes of uneven sensitivity, and rings produced by dust particles in the camera (see the raw images in Fig. \ref{fig: unprojected} panels 1 and 3).
It also would remove the non-variable part of the aurora, which we are therefore not able to detect.
This includes permanent auroral structures fixed in local time.
Without background subtraction non-variable aurora usually also can not be detected because it is indistinguishable from the stray light.
Sometimes images containing aurora had to be used for the background subtraction.
This resulted in oversubtraction and produced permanent dark spots in the resulting movies.
}

\cora{The second noise-reduction technique was the removal of bad detector pixels and cosmic ray hits.
We did this by automatic selection of single pixels that are significantly brighter than any neighboring pixels and replacing them with the average brightness of the neighboring pixels.
This enhances the aurora because the aurora is diffuse and rarely has one-pixel structures in it.
Such removal of bad pixels is important while estimating average auroral brightness, to which they could contribute substantially.
}

On the day shown in Fig. \ref{fig: unprojected}, five night-side images were obtained, two of which show auroras.
They were observed from nearly the same point in space, while Saturn rotated in front of the spacecraft to reveal different longitudes.
\cora{The three images not containing aurora were averaged as a "background image", which was then subtracted from the images in Fig. \ref{fig: unprojected}. 
As will be discussed later, after filtering out the local time effects, nearly-corotating auroral structures are seen in the visible movies, which makes it useful to map the aurora in standard System III longitude.
System III coordinates assume planetocentric latitude and west longitude with the 10.656222-hour SKR period measured by Voyager \citep{desch81}.} 
Details of these observations, and all the other auroral detections up to March 2014 can be found in Table \ref{tab: data}.

\subsection{Auroral Brightness}

\cord{Only the brightest aurora on Saturn (mainly in the main auroral oval) can be detected by the Cassini camera. 
The broadband (250-1000 nm) visible brightness of Saturn's aurora detections ranges from few kilo-Rayleigh (kR) to about 100 kR. 
The tens-of-kR brightness of aurora  in Fig. \ref{fig: unprojected} observation can be judged with the help of the scale bar on the right.
This is less than the 1000-kR typical brightness of Jupiter's visible aurora \citep{vasavada99}.
\core{UV aurora on Saturn (tens of kR) is also about two orders of magnitude fainter than UV aurora on Jupiter \citep{gerard04}.}
}

\subsection{Auroral Latitudes}
\label{sec:auroral_latitude}

In Fig. \ref{fig: unprojected}, two compact auroral spots can be seen at latitude $\sim$ 75{\deg} and longitude $\sim$ 290{\deg} in the first image, both in the raw image shown in Panel 1, and after background subtraction in Panel 2.
The other image shows a faint auroral arc, again at latitude $\sim$ 75\deg.
The arc spans a large range of longitudes and can only be detected after background subtraction, \ie, it can be seen in Panel 4 but not in Panel 3.
\cora{In most of the visible observations, aurora only can be detected after background subtraction.}

\cora{Supplementary Movie \ref{fig: movie199_12} gives a typical example of the auroral fine-structure features that nearly corotate with Saturn.
The features in the movie approximately follow System III longitudes marked on Saturn's "surface".
\core{The main auroral arc at the limb moves from $\sim$-70{\deg} latitude to  $\sim$-75{\deg} latitude 
as time proceeds from 12:33 to 15:00.}
After that  \core{time} aurora becomes faint and non-detecatble.
At about 22:30 the arc appears again. 
Accordingly, the structure observed at 12:33 is observed again about one Saturn's rotation later (at 22:30).
\core{Again, the main auroral arc moves from $\sim$-70{\deg} latitude to  $\sim$-72{\deg} latitude as as time proceeds from 22:30 to 23:48.}
This would make it appear as a spiral on the System III \core{coordinate} map \core{projection combined} from the consecutive images.}
\corb{Similar structures reappear on consecutive Saturn's rotations in several other movies (see Section  \ref{sec:timing}).
\core{The 1000-km-scale wavy structures of the main oval in Supplementary Movie \ref{fig: movie199_12} visibly corotate with Saturn.
This is also true for the similar structures in the movies to be discussed in Section  \ref{sec:timing}.}
Accordingly, we conclude that we observe the night side section of the spiral-shaped auroral oval that nearly corotates with Saturn.}

\begin{figure}[htbp]

    \resizebox{!}{2.6in}{\includegraphics{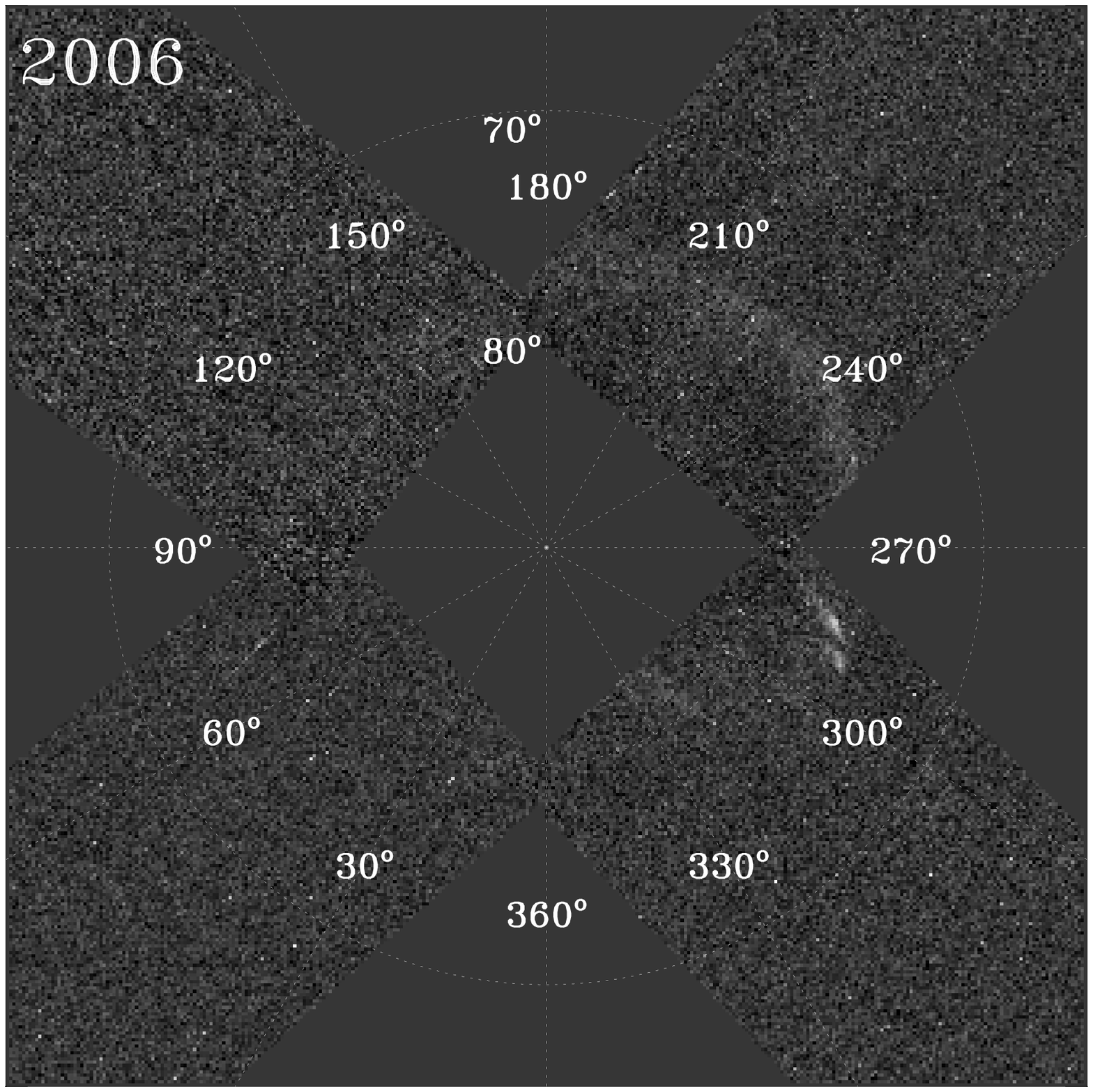}}
    \resizebox{!}{2.6in}{\includegraphics{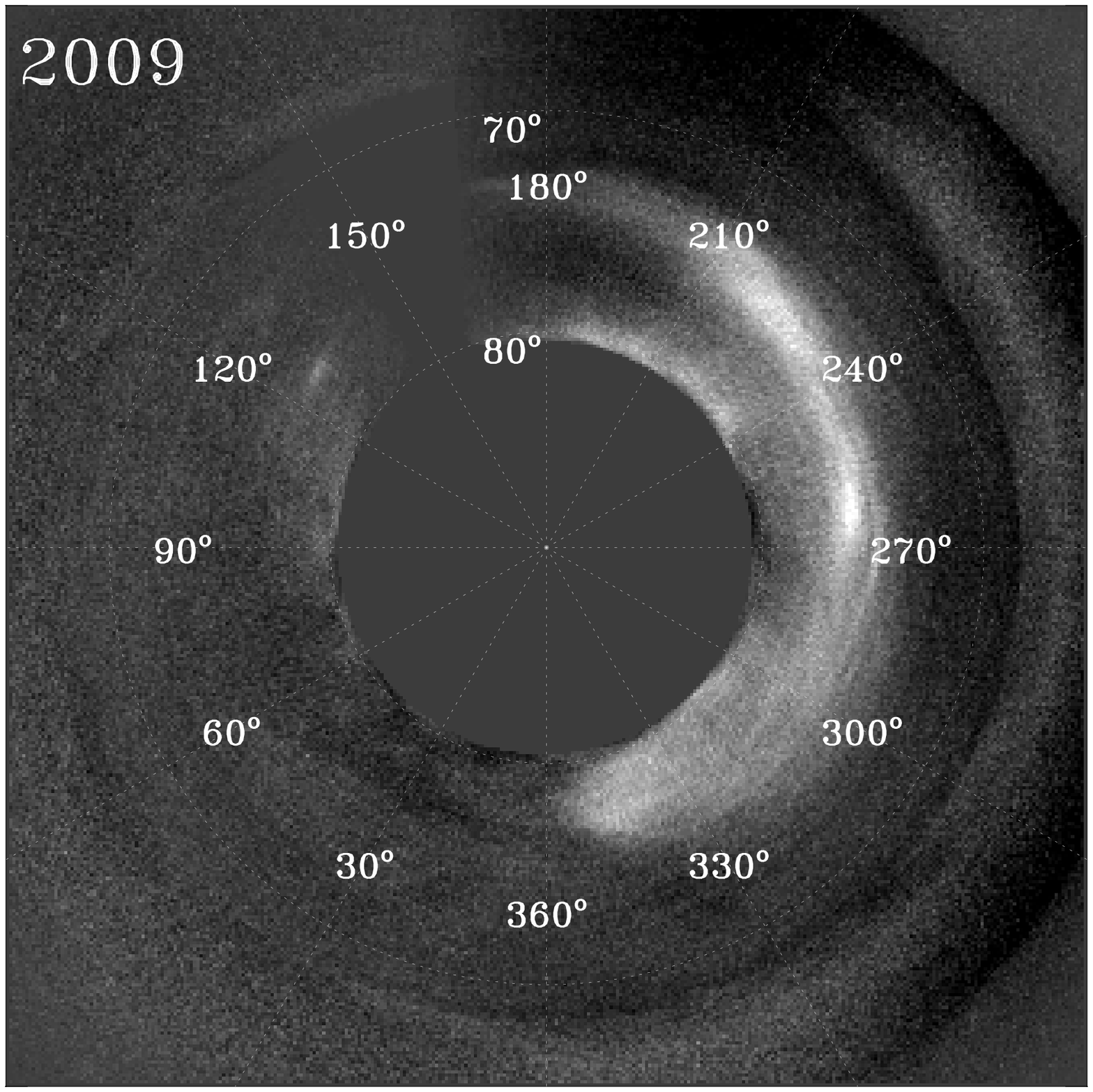}}
    
    \resizebox{!}{2.6in}{\includegraphics{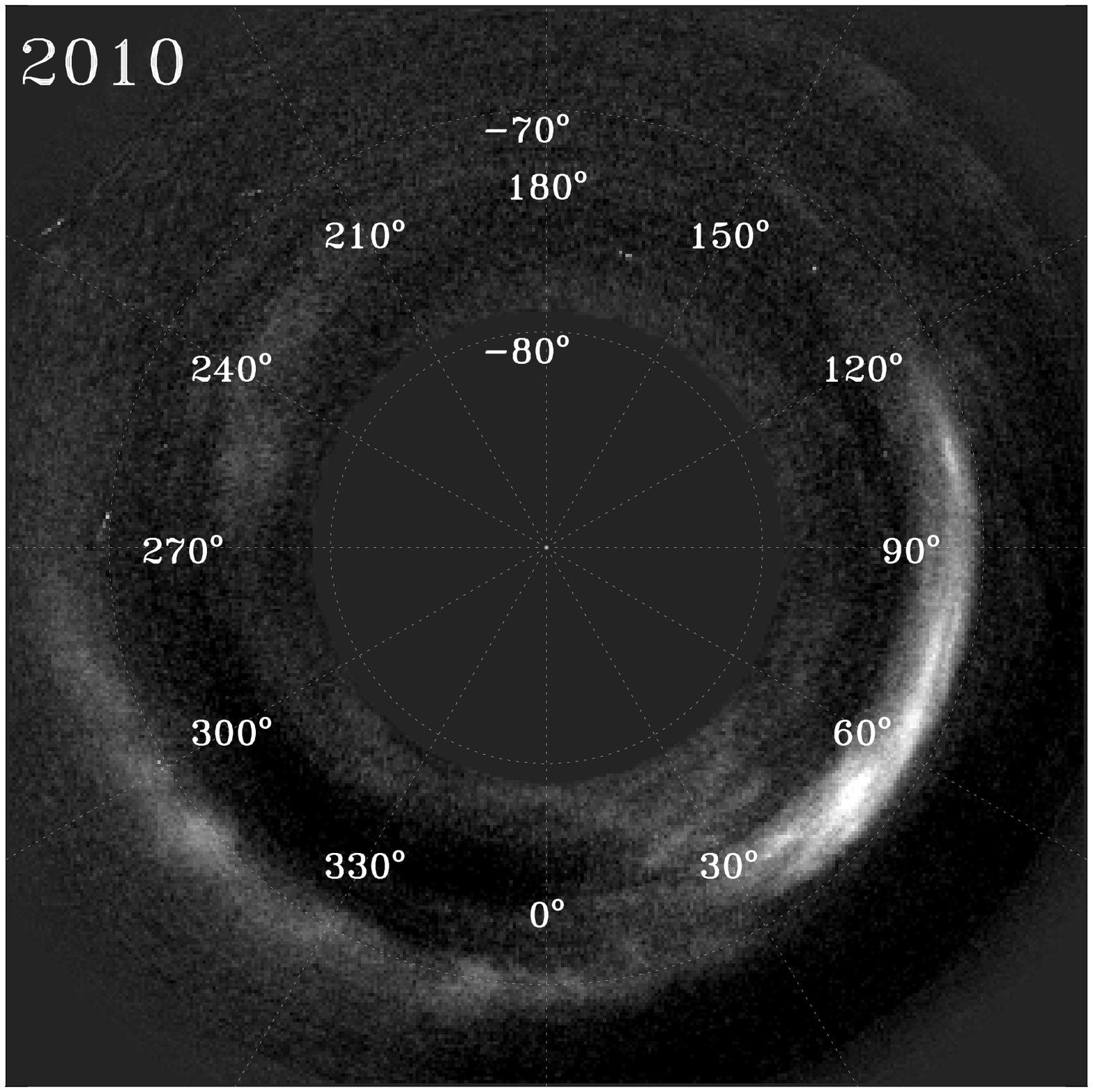}}
    \resizebox{!}{2.6in}{\includegraphics{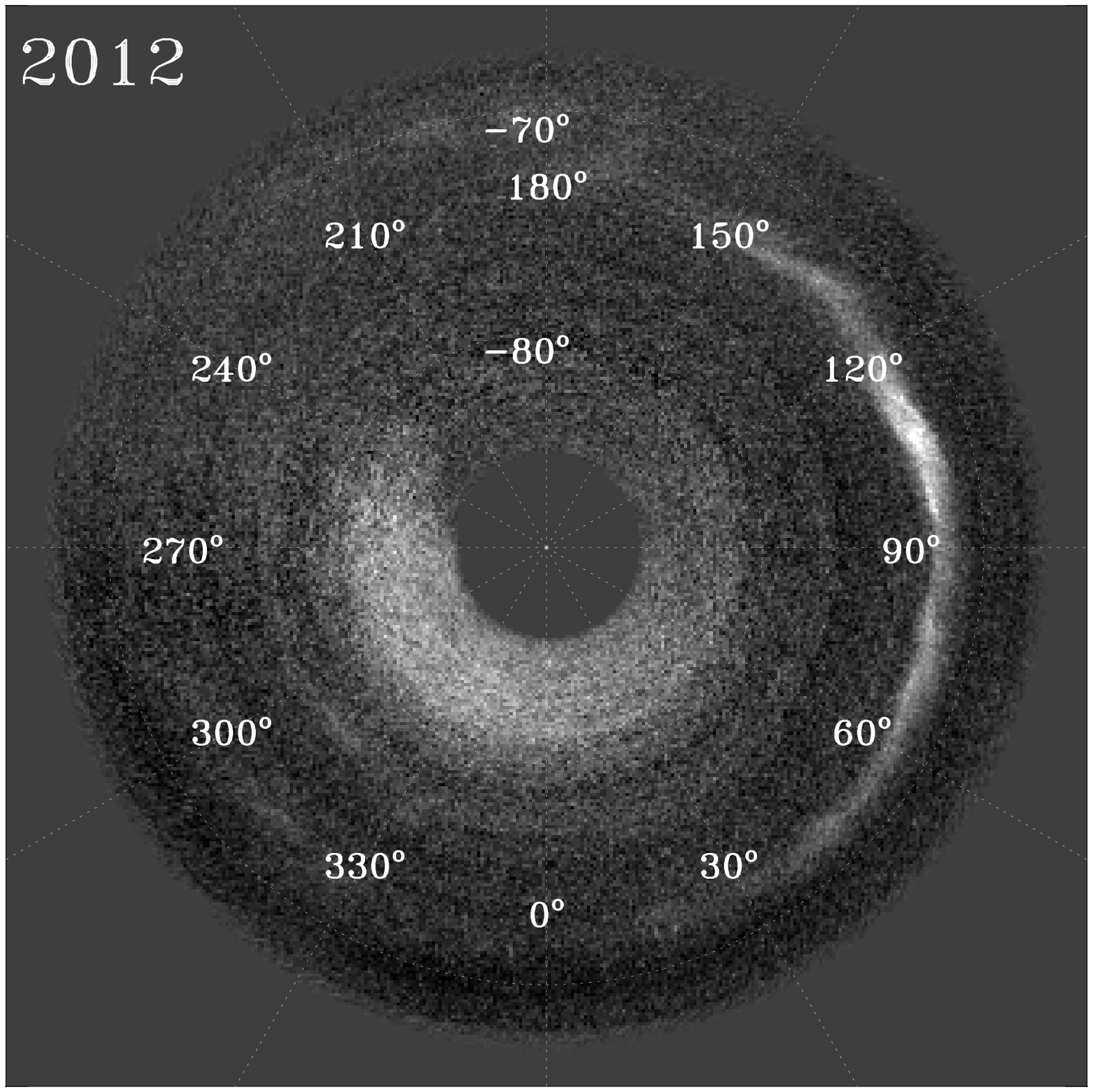}}
    \caption{
\cord{System III \core{coordinate polar azimuthally-projected} maps} of the aurora combined as a mosaic from individual images on the night side of Saturn.
Each image was taken in broadband filter (CL1+CL2) spanning the entire range of visible wavelengths  250 to 1000 nm \citep{porco04}. 
The coordinate grid overlays the maps.
Panels labeled 2006 and 2009 show northern aurora on July 6 (DOY 197) 2006 and Oct. 5 (DOY 278) 2009.
Panels labeled 2010 and 2012 show southern aurora on June 26 (DOY 177) 2010 and July 6 (DOY 197) 2012.
The darkest longitude bands may result from background subtraction, because some of the aurora brightness may be subtracted from the images as longitudinally-averaged background (see also Fig. \ref{fig: unprojected}).
The bright \cora{features} are aurora.
} 
    \label{fig: maps} 
\end{figure}

Figure \ref{fig: maps} shows polar maps combined from nightside images of aurora. 
\renewcommand{\baselinestretch}{1.}
\renewcommand{\baselinestretch}{2.} 
\cora{The northern aurora is observed in 2006 and 2009, and the southern aurora is observed in 2010 and 2012.
}
The main auroral arcs \core{in our images are located between 70{\deg} and 80{\deg} latitude north and between 65{\deg} and 80{\deg} latitude south (as seen in Fig. \ref{fig: maps} and also in the images from Table \ref{tab: data}).}
The 2006 map is created from 5 images, each spanning $\sim$2 hours in local time near midnight, which is equivalent to  $\sim$80{\deg}-wide sector in longitude (see Fig. \ref{fig: unprojected}).
 \cora{Each of the 2009, 2010, and 2012} maps \cora{in Fig. \ref{fig: maps}} is created from a few hundred image segments, each \cora{segment} spanning  $\sim$0.5 hour in local time near midnight.
\cora{Such small range in local time minimizes the effect of local time on auroral latitude (which is a substantial effect, as will be discussed later).}

\cord{Maps for 2009, 2010,  and 2012 in Fig. \ref{fig: maps} and subsequent figures in this paper are small subsets of the data in the movies. 
In the figures it is hard to distinguish (1) features corotating with Saturn from (2) features fixed in local time and (3) other time variabiity. Corotation is obvious in the movies, however, because each movie frame spans $\sim$80 degrees of longitude and there are hundreds of frames in the 9-12-hour intervals (see Table \ref{tab: data}).}

\corb{The typical range of auroral latitudes, as can be seen in Fig. \ref{fig: maps}, is 70{\deg}-80{\deg} South \core{or} North.}
The latitudes  of the visible-wavelength main oval are consistent with the 10{\deg}-25{\deg} co-latitudes of the main oval  observed in IR \citep{badman11} and UV \citep{carbary12,lamy13}.
\cite{melin11} show in simultaneous UV and IR observations that some IR auroral features are also observed in UV, though others are not.
\cite{melin14}, this issue, show colocation of UV, IR, and visible aurora observed by Cassini in spring 2013.
\cord{This suggests that similar processes are responsible for aurora observed in all three wavelengths}.


\corb{It is interesting to estimate what parts of the magnetosphere map to the auroral latitudes. 
For spin-alingned magnetic dipole the co-latitude $\theta_0$ of the footprint of the field line, which crosses magnetic equator at distance $a$ is the following \citep[from][]{goldreich07}.}

\corb{$\sin \theta_0 = (R_S/a)^{1/2}$},

\corb{where $R_S$ is Saturn's radius. 
Accordingly, latitudes of 70{\deg} and 80{\deg} ($\theta_0$=20{\deg} and $\theta_0$=10{\deg}) map to $a \approx$8$R_s$ and $a \approx$33$R_s$.
}

\corb{\core{A more realistic model includes magnetosphere interacting with the interplanetary magnetic field \citep{belenkaya11,belenkaya14}. 
It} is restricted by magnetic field measurements by Cassini.
The model projects the open-closed field line boundary at the night side to latitudes 75{\deg}-80{\deg} for most of the \core{magnetic field measurements}.
The ultraviolet dayside aurora, \cord{observed simultaneously with the magnetic field measurements, is located} predominantly equatorward from the open-closed field line boundary footprint.
Given the 65{\deg}-80{\deg} latitudes of \core{the main auroral oval} in our visible observations, this may also be the case.
However, latitudes vary substantially \cord{with time in both our auroral observations and the modeled} open-closed field line boundary.
\core{Because there are no simultaneous magnetospheric observations, the exact position of aurora relative to the open-closed field boundary in our observations is  unknown.}}

\subsection{Spiral Morphology}

In the \core{rotating System III polar azimuthal projections} in Fig. \ref{fig: maps} the auroral arcs spiral toward the pole in the westward direction, \ie, \cora{in the direction of increasing longitude, which is clockwise in North 2009 map, and counterclockwise in South  2010 and 2012 map}.
\core{This can be clearly seen in 2009 and 2012 maps.}
\cora{\core{The same spiraling direction was discussed in Section \ref{sec:auroral_latitude} for Supplementary Movie \ref{fig: movie199_12}, which was taken 2 days later than the 2012 map in Fig. \ref{fig: maps}.}
Interestingly, this spiraling direction would also be expected from the azimuthally-localized centrifugal plasma outflow in the magnetospheric equator, or  "plasma tongue" \citep{goldreich07}.
The plasma in the tongue flows away from Saturn and nearly corotates with Saturn.}
Its rotation slows down with distance from Saturn.
\corb{The tongue itself co-rotates with Saturn, but it forms a spiral arm trailing behind (to the west of) the source in Saturn's equatorial plane}.
\cora{If we assume that the tongue projects along the magnetic field lines to the polar aurora, the more distant parts would map closer to the pole.
This gives the same spiraling direction as the observed aurora in Supplementary Movie \ref{fig: movie199_12}, Fig. \ref{fig: maps} and other visible movies}.
\corb{This model also explains the corotation of the spirals with Saturn.}

\cora{Spirals are also observed in ultraviolet images of Saturn's aurora, \eg, by Hubble \citep{gerard04, grodent05} and Cassini \citep{pryor11,radioti11}.
\corb{They spiral \core{to the pole clockwise at the North, and counterclockwise at the South, which is} the same direction as in our images.}
These ultraviolet spirals were explained by reconnection of the magnetosphere with the solar wind \citep{cowley05}.
This explanation assumes that, like on Earth, the oval maps to the open-closed field line boundary.
Our visible aurora observations define corotation with SKR within 1-2\%  (as will be discussed in Section \ref{sec:timing})}.
\cord{Previous studies, e.g.,  \citep{grodent05}, report separate auroral features on the day side changing their speed from 70\% to 20\% of corotation rate.
The 100\%} corotation in ultraviolet aurora was detected only recently \citep{lamy13}.
Accordingly, the model explaining ultraviolet spirals by \cite{cowley05} did not attempt to explain corotation. 
\core{Another model 
proposes coupling of the polar neutral atmosphere with the open field lines, producing corotation poleward from the open-closed field line boundary \citep{southwood14}}

\corb{ \cite{nichols08} reported \cord{near-planetary period oscillations} in the appearance of the ultraviolet auroral oval.
They approximated the moving dayside part of the oval as a circle with time-dependent radius and the center oscillating with Saturn's period.
\core{Circle fitted by  \cite{nichols08} do not contradict with the spiral we observe.
The dayside part of a near-corotating spiral-shaped main auroral arc would also give a good fit to such an oscillating circle if the spiral shape is not well resolved, as in the UV images in  \cite{nichols08}.}
}

\cora{Some observations suggest \corb{ aurora mapping to the open-closed field line boundary or to the inner magnetosphere near that boundary}. 
\cite{bunce08} proposed that the boundary is mapping directly to aurora.
They observed the sheet of upward field-aligned current, potentially associated with aurora, at the same location as the open-closed field line boundary when Cassini was crossing field lines mapping to aurora.
However, \cite{talboys09} places in situ observations of the upward field-aligned current on closed field lines just equatorward of the open-closed field line boundary.
\cite{belenkaya11, belenkaya14} track interplanetary magnetic field measured by Cassini in situ to the ultraviolet aurora location.
They conclude that aurora extends  equatorward from the open-closed field line boundary to the outer boundary of the equatorial ring current.}

\corb{The plasma tongue by \cite{goldreich07} may also produce spiraling aurora when \core{the plasma in the tongue does not directly project along the field lines to the main oval}, but when \core{plasma tongue} distorts the magnetic field and, accordingly, the open-closed field line boundary.
Such magnetic field distortions were modeled by \cite{nichols08}. 
The tongue was modeled as a half-circle rather than spiral.
It pulled the nearby open-closed field line boundary to lower latitudes.
The spiral shape of the tongue would create spiral-shape distortion in the boundary, as observed in aurora.} 

\cora{The spiral dynamics reported here may be used to test which magnetospheric region 
maps to the main auroral oval.
Such a study needs additional magnetospheric modeling and is outside the scope of this paper.
}

In our 2010 observation, the aurora formed a double arc \cord{in the main oval} (longitudes 270{\deg} through 30{\deg}  in the 2010 map and in Supplementary Movie \ref{fig: movie177_10}).
\corb{Double arcs \cord{forming the main oval} are sometimes seen in ultraviolet \citep{grodent05, stallard08, radioti11}.
\core{Some of the ultraviolet images show \cord{the main oval as a spiral going around the pole and continuing as the second} arc closer to the pole}.
Other ultraviolet images show double arcs \core{in small azimuthal segments, which are not connected into a spiral going around the pole}.
In  Supplementary Movie \ref{fig: movie177_10} the entire oval structure \cord{around the pole} can not be tracked to determine whether the double arc represents the two edges of a continuous spiral.
The movie is long enough to cover a whole oval's rotation, and samples all longitudes as they pass by midnight sector.
However some parts of the oval are faint and undetectable, not allowing one to tell if the two arcs connect into a continuous spiral.
}

\cora{Spirals that nearly corotate with Saturn were observed in ion data from CAPS (the Cassini Plasma Spectrometer).
\corb{They spiral in the same direction as in UV and visible aurora.}
They project to the magnetospheric equator at 10-50 $R_S$.
\cite{burch09} \core{propose} that the observed dense nearly corotating "plasma cam" inside 10  $R_S$ is responsible for plasma loading to the spirals.
Two possible origins of the cam were proposed: closed field line reconnection; or plasma convection inside magnetosphere, in which case the cam is the extension of plasma tongue from \cite{goldreich07}.
}

\subsection{Small-Scale Structures}

Interesting linear features can be seen in the 2010 map in  Fig. \ref{fig: maps} at 30{\deg} longitude poleward from the brightest arc (latitudes 75{\deg}-80{\deg}).
In Supplementary Movie \ref{fig: movie177_10}  they appear as rather faint curves with lifetimes of at least 10 min. 
These curves project to the polar map as straight lines.
At least two bands at latitudes $\sim$75{\deg} and $\sim$78{\deg} are aligned perpendicular to the 30{\deg} meridian.
The mechanism forming these bands is unclear.
\core{Linear polar arcs approximately perpendicular to the meridian were also observed in UV by Cassini \citep{radioti14}.
Magnetotail reconnection was discussed as a possible cause of them.
Those arcs, however, were seen in the dusk sector of the auroral oval, while our visible bands are near the midnight.}
A possible terrestrial analog to the visible bands reported here may be  the  polar rain aurora which forms bands perpendicular to the midnight meridian \citep{zhang07}.
Polar rain aurora is formed by solar wind electrons precipitating inside the auroral oval along the open field lines.
\core{More evidence is needed to determine whether the polar rain is related to the auroral bands on Saturn.}

\begin{figure}[htbp]
    \resizebox{!}{2.6in}{\includegraphics{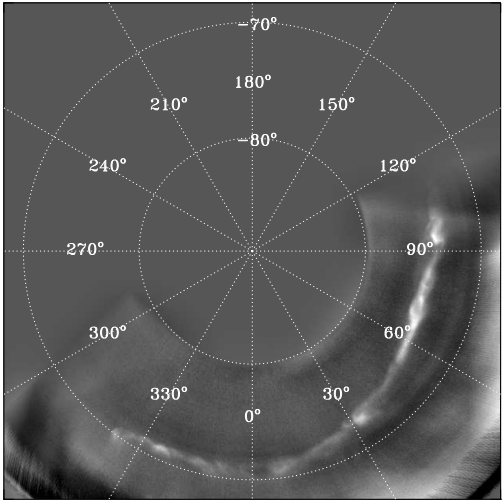}}
    \caption{
Detailed auroral structure of southern aurora on  Nov 24 (DOY 329) 2012.
The \core{System III coordinate polar} map is produced in the same way as in Fig. \ref{fig: maps}.
The map was high pass filtered to reduce the effects of stray light.
The images were obtained near dawn ($\sim$3 to 5 hours local time).
Structures at latitudes lower than 70{\deg} are stray light contamination.
The bright arc at latitude 70-75{\deg} is aurora.
} 
    \label{fig: fine_structure} 
\end{figure}

Figure \ref{fig: fine_structure} shows the structure of the southern auroral arc in the movie taken on  November 24 2012.
\renewcommand{\baselinestretch}{1.} 
\renewcommand{\baselinestretch}{2.} 
The map is made from Supplementary Movie \ref{fig: movie329_12}, which has the highest time resolution of $\sim$1 minute per frame.
Bright clumps \cord{measuring about 1{\deg} latitude (1000 km)} are seen along the arc in Fig. \ref{fig: fine_structure}.
The \ref{fig: movie329_12} movie shows that these clumps nearly corotate with Saturn.
\cora{The clumps change their speed on timescales of tens of minutes, sometimes superrotating and sometimes lagging behind the coordinate grid.
They appear as} anticlockwise  vortices extending poleward from the main arc.
When they extend poleward, they brighten and accelerate in prograde direction.
\corb{This is consistent with shear flow with faster rotation at the polar side of the oval.
This may indicate an open-closed field boundary.
In this case the polar cup near-corotates with Saturn's neutral atmosphere, but the outer magnetosphere just eauatorward from the boundary subcorotates \citep{southwood14}.}

\core{
The mechanism producing our auroral vortices may be related to a magnetospheric vortex observed in situ by Cassini crossing the dayside open-closed field line boundary \citep{masters10}.
The magnetospheric vortex was observed on the planet side of the boundary, and was tentatively explained by Kelvin-Helmholtz instability in the sheared flow at the boundary.}

\core{Similar-size (few thousands of km) spots were observed in 500-km-resolution Cassini UV observations \citep{grodent11}.
The UV observations produce "pseudoimages" by slewing the UVIS instrument slit across the image for about half an hour, which confuses spatial and temporal effects.
As is apparent from Supplementary Movie \ref{fig: movie329_12}, the aurora shows brightness waves and  bright clump accelerations on the half-an-hour timescale, which probably contributed to the appearance of the main auroral arc in UV pseudoimages as "fragmented into
small substructures" \citep{grodent11}.
At 31 km/pixel resolution of Supplementary Movie \ref{fig: movie329_12} the $\sim$500-km-wide visible-wavelength main auroral oval appears to be continuous between the brighter clumps.}

\subsection{Auroral Latitude Versus Local Time}

\begin{figure}[htbp]
     \resizebox{!}{1.8in}{\includegraphics{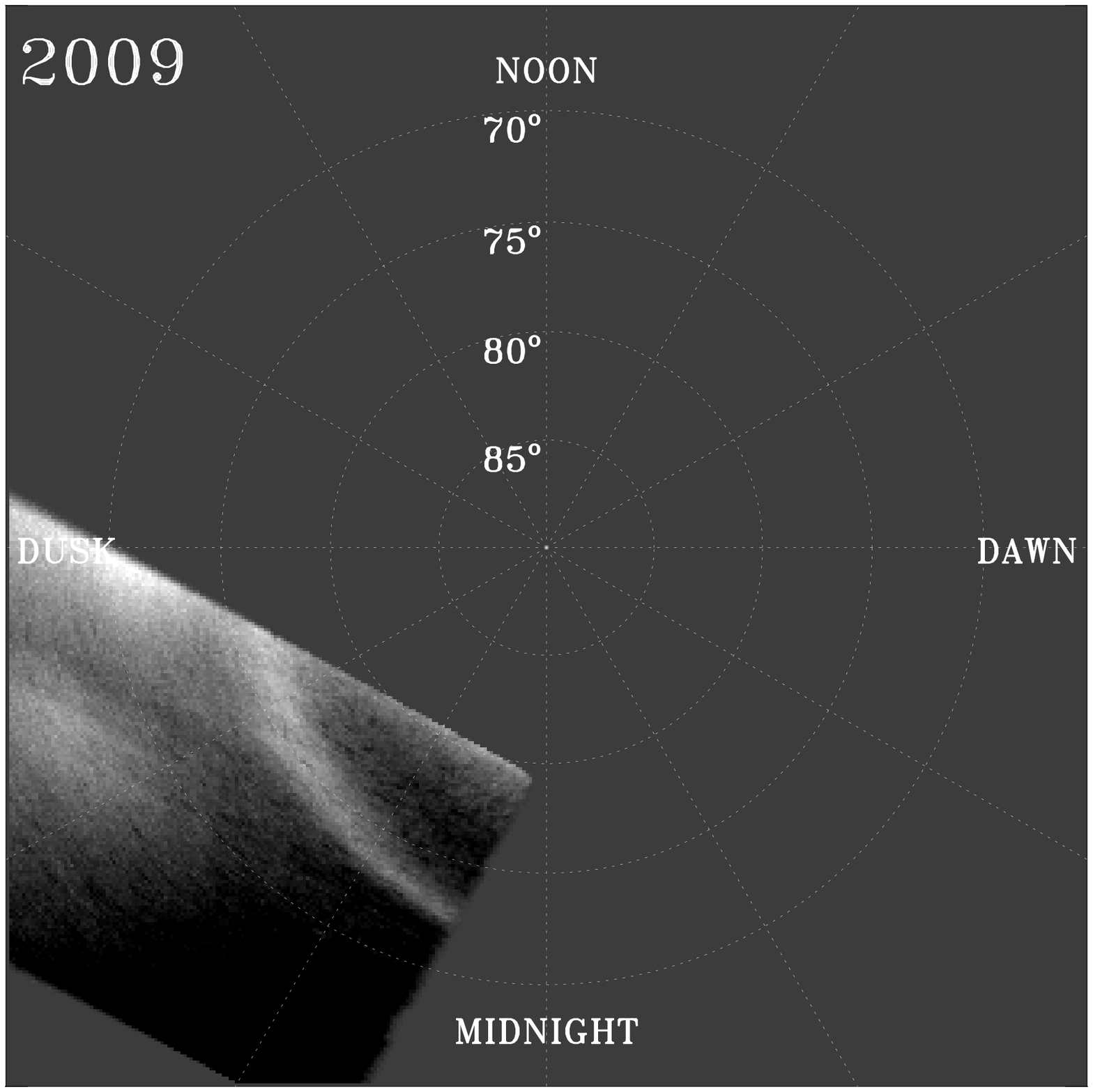}}
    \resizebox{!}{1.8in}{\includegraphics{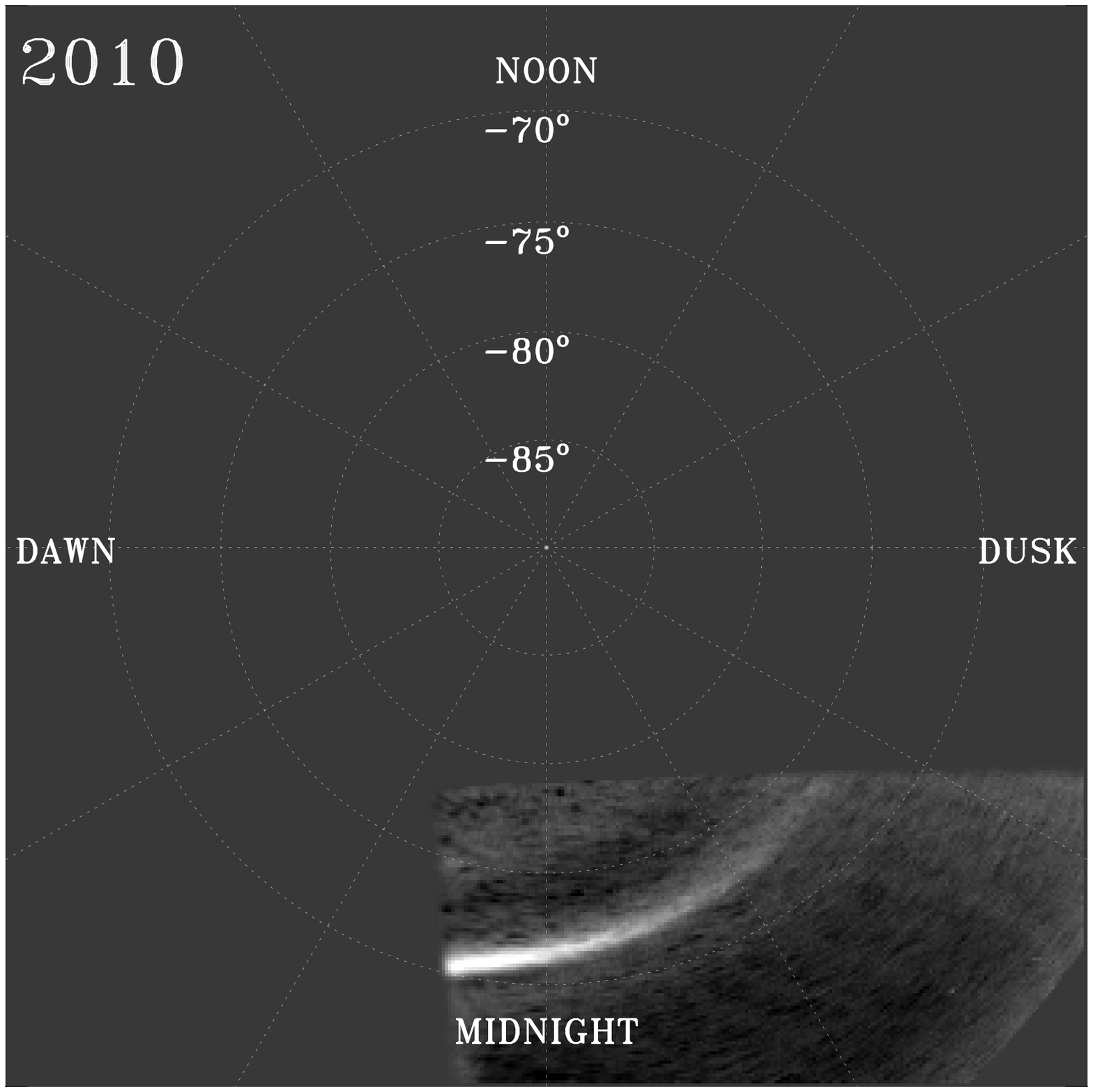}}
    \resizebox{!}{1.8in}{\includegraphics{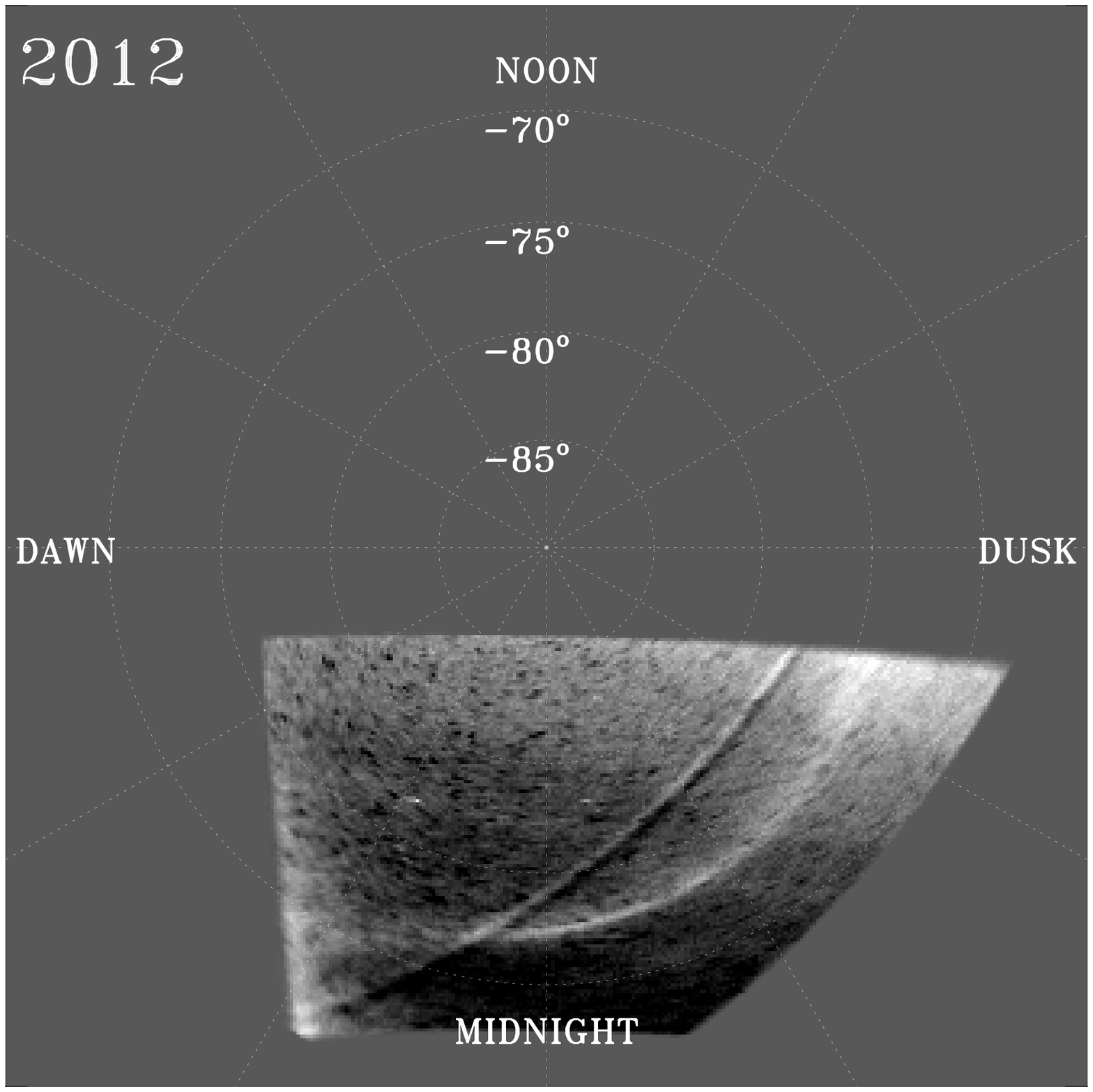}}
    \caption{
Maps of the auroral location versus local time.
\cord{The projections of individual movie frames taken at different times are \core{mapped as functions of local time and then averaged together for each of the  movies also used for Fig. \ref{fig: maps}.}}
The 2009 map averages aurora over 1h 40m from longitudes 180{\deg} through 270{\deg} (the bright arc in the "2009" panel of Fig. \ref{fig: maps}). 
The 2012 map averages aurora over 6h 26m from longitudes 240{\deg} through 360{\deg} and through 210{\deg} (nearly a full circle in the "2012" panel of Fig. \ref{fig: maps}). 
The grid of latitudes and local times overlays the maps.
Note that dawn and dusk in the 2009 North pole map are mapped opposite to the other maps showing the South pole.
} 
    \label{fig: local_time} 
\end{figure}

Figure \ref{fig: local_time} shows the aurora location versus local time. 
\renewcommand{\baselinestretch}{1.} 
\renewcommand{\baselinestretch}{2.} 
\cord{The maps are the average of the instantaneous local-time maps of individual movie frames.
Only the brightest segments of the rotating auroral oval contribute substantially to the bright arcs in the Fig. \ref{fig: local_time} average images.
The bright segments, 90-150{\deg}-wide in longitude, can be seen in the rotating frame in Fig. \ref{fig: maps}.
The time-averaging is essential to avoid confusion of the local time main oval latitude variation with the azimuthal structure rotating with System III \core{coordinates} (main oval latitude varies with azimuth in Fig. \ref{fig: maps}).
Both local time and and azimuthal dependence of main oval's latitude contribute to an instantaneous local-time map of individual movie frame.
In the averaged maps the rotating structures average out and only variation of main oval's latitude with local time is seen.}
\corb{The range of local times is restricted by Cassini's viewing position \cord{(at the night side of Saturn)} and by the camera's field of view during each auroral movie}.

At midnight, the aurora appears at 2-4{\deg} lower latitude than at the dusk.
\core{This value varies from observation to observation}.
\cord{In the movies this effect is seen as the \core{small-scale structures in the} main oval} 
\corb{
moving to lower latitudes as they \cord{rotate from the dusk to} midnight.
\cord{\core{This effect is seen in the movies at all longitudes of the rotating main oval, though only the bright part of the oval in each movie contributes to Fig. \ref{fig: maps} produced by movie frame averaging.}
\core{Consistent displacement of the auroral arc for all longitudes of the rotating oval to lower latitudes near midnight} may indicate that the center of \cord{corotation} is displaced toward midnight or dawn by several degrees latitude.}
This is the same sense of displacement as the 1.6{\deg} anti-sunward displacement of the auroral oval in IR and UV observations \citep{badman11}.
\cora{Also, it is consistent with \core{1-2{\deg} dawn-midnight-directed displacement of the fitted rotation center of the aurora observed in UV by Cassini \citep{badman06,carbary12} }.
 \corb{\core{Also, it is consistent} with 1.8-2.2 degree offset to 3-4 h LT for the center of the oscillating circles fitted to the UV auroral oval by \cite{nichols08}.}}
\cord{Finally,} the displacement is consistent with the 2-4{\deg} anti-sunward displacement of modeled projection of open-closed field line boundary produced by the solar wind distorting the magnetosphere \citep{belenkaya11,belenkaya14}.}

\section{Temporal Variations and Periodicities.}
\label{sec:timing}

\subsection{10-minute Brightenings}

The temporal variations of the aurora can best be viewed in the Supplementary movies.
The aurora can vary dramatically over time, changing from a near quiescent state to a bright auroral arc \cord{or to a separate spot and back to quiescent state} over timescales as short as 10 min (see Supplementary Movie \ref{fig: movie280_09}, around time 23h00m).
Such brightenings repeat about every hour $\pm$ 10 min, as is seen in \cord{about half a dozen of the visible} auroral movies \cord{to be shown as Supplementary Movies in this paper}. 

\subsection{Corotation with Saturn}

The wavy shape of the auroral arc in Supplementary Movie \ref{fig: movie280_09} stays the same during these brightenings and approximately corotates with Saturn (it follows the \cora{System III} coordinate grid in the movie).
Fainter and less variable auroral structures also appear to corotate with Saturn.

This corotation raises the obvious question of whether aurora can \cora{provide some information about} 
\cord{the rotation periods of Saturn's atmosphere and magnetosphere.}
\cora{The disturbances of} Saturn's magnetosphere, which the aurora represents, 
\corb{are expected to be connected to Saturn's upper atmosphere, which is influenced by the deeper atmosphere} \citep{goldreich07,gurnett07,cowley13,fischer14}.

Two long sets of auroral movies  were taken in 2009 and 2012.
These sets of movies were interrupted by gaps in observations and cover 8 and 11 Saturn rotations for 2009 and 2012, respectively. 
\cora{We assumed different 'auroral' rotation periods to test which period best lines up features \cord{that appear similar} from rotation to rotation.}
Supplementary Movies \ref{fig: parallel_movie281_09} and \ref{fig: parallel_movie199_12} \cora{show parts of the movies from different rotations for the best-fit cases}.
\cord{By lining up movies on different rotations} the motion of the auroral structures can be compared. 
We tested different 
\cora{rotation} rates between 10.4 and 11 hours \cord{and produced the movies like Supplementary Movies \ref{fig: parallel_movie281_09} and \ref{fig: parallel_movie199_12} for each rotation rate tested}.
\cora{For the best-fit cases in} Supplementary Movies \ref{fig: parallel_movie281_09} and \ref{fig: parallel_movie199_12}
the auroral patterns on different rotations \cord{visually} resemble each other.

We used three criteria to determine such resemblance.
First, in Fig. \ref{fig: maps}, the \cora{bright} auroral arc in the \cora{2009 and 2012} maps form spiral \cora{segments}.
The low-latitude end of \cord{these spiral segments} can be seen in Supplementary Movies \ref{fig: parallel_movie281_09} and \ref{fig: parallel_movie199_12} when the aurora at the limb reaches its lowest latitudes.
\cord{This happens when} longitude $\sim$200{\deg}  \cord{crosses the limb} in \ref{fig: parallel_movie281_09} movie and \cord{when longitude} $\sim$30-60{\deg}  \cord{crosses the limb} in \ref{fig: parallel_movie199_12} movie. 
We will call this \cord{location in the rotating frame} the "low-latitude extremum".
Another criterion is the brightest part of the auroral arc.
\cora{It is clear in the movies that one side of the \cord{rotating} auroral oval is always bright, and the other side is dark.
For example, in Supplementary Movies \ref{fig: parallel_movie281_09} at longitudes $\sim$180-360{\deg} \cord{bright} aurora is seen on all rotations, but at longitudes $\sim$0-120{\deg} aurora is \cord{faint and} not detectable on any rotations.
However, the brightness maximum \cord{is broad} in longitude \cord{and its longitudinal location is less certain than} the location} of the low-latitude extremum.
The third criterion is the part of the auroral arc which brightens suddenly on a one-hour period, \cora{unlike the \cord{steadily-glowing} other parts of the oval.
This "blinking" part of the oval can only be seen in the movies.
It occupies longitudes} 180{\deg}-240{\deg} in \ref{fig: parallel_movie281_09} movie and 0{\deg}-50{\deg} in \ref{fig: parallel_movie199_12} movie, \cora{respectively,} and gives a moderately good reference frame for matching aurora on \cord{different rotations}.

\cora{\corb{The nearly-corotating aurora changes brightness on shorter timescales than Saturn's rotation \cord{period}}.
Small-scale \cord{(1000 km)} auroral features change speed from superrotation to subrotation \corb{on timescales of hours} (see Supplementary Movie \ref{fig: movie329_12}).
It is possible that they change shape or drift substantially in longitude \cord{over the 10-hour Saturn's day}.
Because of \cord{these two reasons}, we did not use the small-scale features to determine an auroral rotation period.
Instead, we used the \cord{large-scale} auroral structure described by the three criteria above.
\cord{This structure} persists from rotation to rotation.}
Matching aurora with all three criteria gives \cora{the rotation period} for each of the 2009 and 2012 movies.
The best-fit periods are 10.65$\pm$0.15hr (\cora{the same as} the Voyager rotation rate) for 2009 and 10.8$\pm$0.1 hr for 2012.
The uncertainties are from day-to-day variation of aurora.
The error bars are determined from the range of reasonably good-matching movies similar to Supplementary Movies \ref{fig: parallel_movie281_09} and \ref{fig: parallel_movie199_12}, which were produced for various rotation periods.

\begin{figure}[htbp]
    \hspace{.2in}    
    \resizebox{!}{5.1in}{\includegraphics{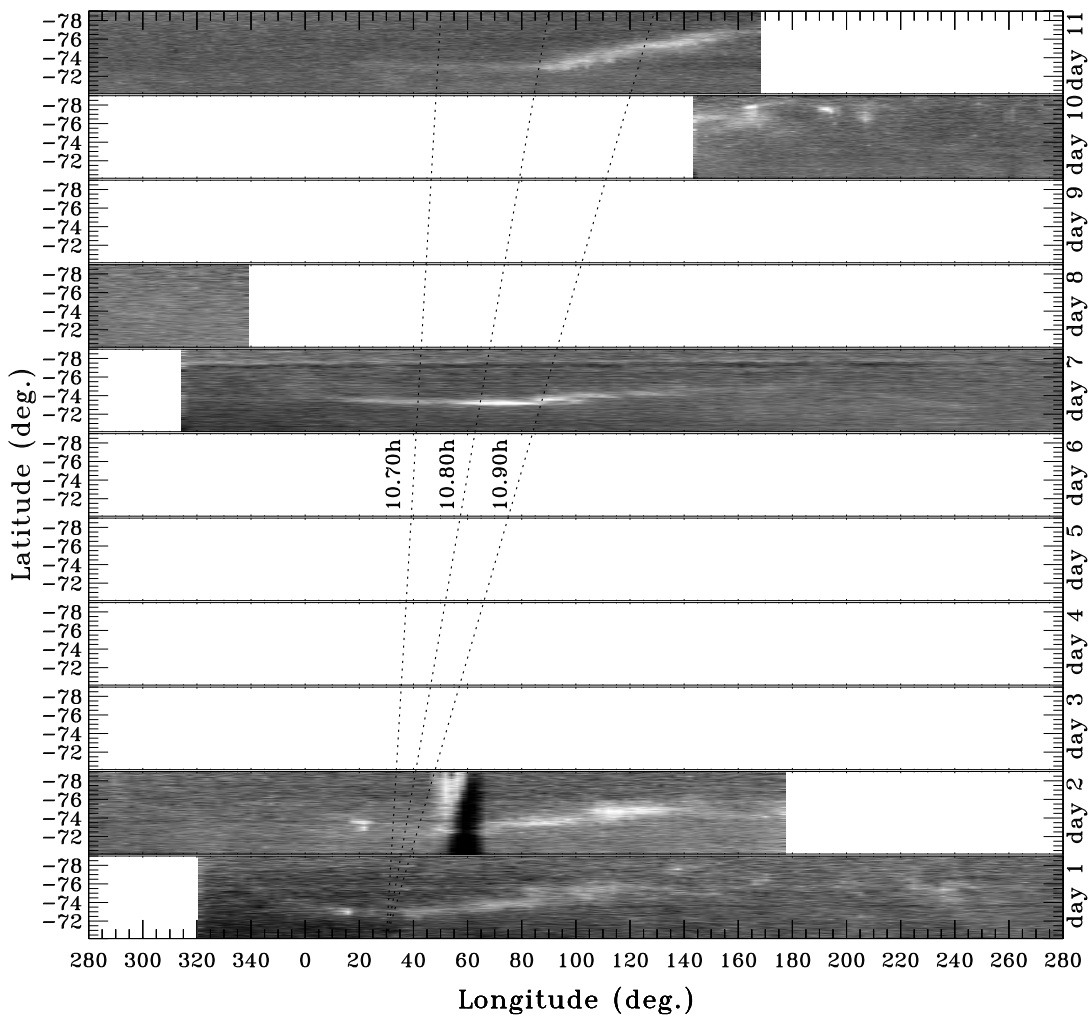}}
   \vspace{0.3in}
    \caption{
Maps of aurora observed on different Saturn rotations on  July 4-8 (DOY 195-199) 2012.
For each movie frame, aurora is mapped in a $\sim$ 5-degree longitude swath around 22h local time.
This  produces a 4-frame overlap at each longitude.
The maps from different movie frames are then averaged as longitudes move through the local time.
The resulting maps from different Saturn days are then stacked as vertical panels, with white space in data gaps.
Panel labels indicate \cord{time in Saturn days}.
West longitudes assume the coordinate system III with Saturn rotation rate of 10.6562 hours.
The time when Cassini observed a particular longitude goes from left to right.
Different aurora rotation periods would cause auroral structures to drift in longitude with time.
Each dotted line indicates the drift corresponding to the rotation period marked next to it.
The prominent feature on day 2 at longitude $\sim$ 320{\deg} is stray light from Saturn's moon Mimas passing across the field of view. 
}
    \label{fig: doy195_12_lat_plot} 
\end{figure}

Another way to demonstrate the rotation period of the auroral features is to map aurora and compare the maps from different Saturn days.
Figure \ref{fig: doy195_12_lat_plot} shows such maps for the 2009 movie set.
\renewcommand{\baselinestretch}{1.} 
\renewcommand{\baselinestretch}{2.}
The maps are poduced from multiple images that were also used for Supplementary Movies \ref{fig: parallel_movie199_12}.
A small range of local times \cora{($\sim$ 10 min LT)} in each image was used for the maps to avoid local time effects (shown in Fig. \ref{fig: local_time}).
This restricted the spatial coverage of the individual images.
Also, vertical aurora structure, including above-the-limb aurora, was ignored in the maps shown in Fig. \ref{fig: doy195_12_lat_plot}. 
\cord{Also, this caused an ambiguity between corotating features and time variations at fixed System III longitude.}
Because of these limitations, the maps miss some structures seen in Supplementary Movie \ref{fig: parallel_movie199_12}.

Some features and periods can be seen in the maps in Fig. \ref{fig: doy195_12_lat_plot}.
The low-latitude extremum appears  at about  40{\deg} on \cord{Saturn's rotation 1 (labeled "day 1" in Fig. \ref{fig: doy195_12_lat_plot})} and drifts \cord{west (right in Fig. \ref{fig: doy195_12_lat_plot}) } on later days.
The brightest part of the auroral arc also moves right from day 1 to day 11.
The drift rates derived from Supplementary Movie \ref{fig: parallel_movie199_12} are shown by the dotted lines in Fig. \ref{fig: doy195_12_lat_plot}.
They connect  the  low-latitude extremum on day 1 with its possible location on the next days, giving a good fit.
The brightest part of the arc also drifts at similar rate.
This confirms the 10.8$\pm$0.1 hr period derived from Supplementary Movie \ref{fig: movie329_12}.
\cora{The \cord{$\pm$0.1 hr error bar on the period corresponds to the left and right} dotted lines in Fig. \ref{fig: doy195_12_lat_plot}.}

\cord{Other features can only be seen in the movies but not in the maps in Fig. \ref{fig: doy195_12_lat_plot}.}
The \cora{blinking} part of auroral arc cannot be recognized in Fig. \ref{fig: doy195_12_lat_plot} because it lacks the combined spatial and temporal coverage of the \ref{fig: parallel_movie199_12} movie.
\corb{For the same reason}, the \cord{the small-scale wavy structures moving with the coordinate grid in Supplementary Movie \ref{fig: parallel_movie199_12} can not be seen in Fig. \ref{fig: doy195_12_lat_plot}.}
 
\begin{figure}[htbp]
    \hspace{.2in}    
    \resizebox{!}{5.1in}{\includegraphics{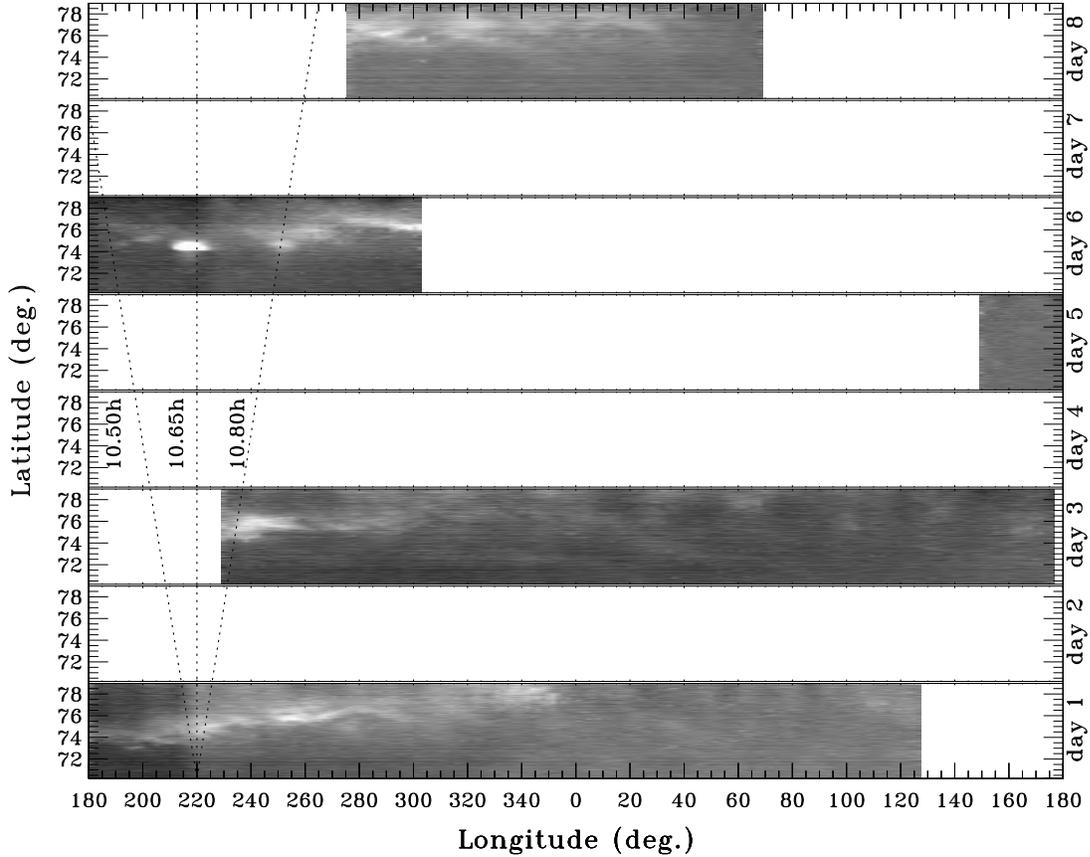}}
   \vspace{0.3in}
    \caption{
Maps of aurora observed on different Saturn's rotations on  Oct. 5-9 (DOY 278-282) 2009 produced similarly to Fig. \ref{fig: doy195_12_lat_plot}.
For each movie frame, aurora is mapped in a $\sim$ 7-degree longitude swath around 11 p.m. local time.
This  produces a 5-frame overlap at each longitude. 
Dotted lines indicate the drift corresponding to the rotation period marked next to it.}
    \label{fig: doy278_09_lat_plot} 
\end{figure}

 Figure \ref{fig: doy278_09_lat_plot} shows maps similar to Fig. \ref{fig: doy195_12_lat_plot} but constructed from the 2009 movies.
\renewcommand{\baselinestretch}{1.} 
\renewcommand{\baselinestretch}{2.}
With  shorter time coverage in 2009 the location of the low-latitude extremum is less clear than in 2012.
However the period of  10.65$\pm$0.15 hr (dotted lines in Fig \ref{fig: doy278_09_lat_plot}) is consistent with both low-latitude extremum and the brightest aurora locations on different days.

\subsection{1-hour Period}

Another period can be seen in Fig \ref{fig: doy278_09_lat_plot}.
It is the $\sim$1-hour period which was already seen between auroral brightenings in the movies. 
Cassini scans  different longitudes \cord{passing by the small local time sector used for Fig. \ref{fig: doy278_09_lat_plot} as time proceeds}.
\cord{In such setup the 10-minute-long brightenings of \cord{longitudinally-extended sectors of the auroral oval} that are $\sim$1-hour apart in time} would appear about 30-40{\deg} longitude apart (Saturn rotates at about 34{\deg} longitude per hour).
This separation can be seen in the brightest auroral spots on day 6 at longitudes $\sim$220{\deg}  and 260{\deg}, also between the bright spots on other days, and, interestingly,  at the virtually aurora-free longitudes 0{\deg} through 180{\deg} on day 3.
A similar period of $\sim$60-80 minutes was observed by Cassini in auroral hiss radio intensity \citep{mitchell14}, in pulses of field aligned energetic electrons and ion conics \citep{mitchell09}, \cora{in the in situ measurements of plasma waves, energetic ions, electrons, and magnetic field \citep{badman12a}, and in ultraviolet aurora \citep{radioti13}. 
 \cite{radioti13} propose that the $\sim$1-hour period in brightening of the dayside ultraviolet auroral oval is induced by magnetic flux tube reconnections.
 }

\subsection{Relation of Aurora Corotation to PPO}

It is interesting to compare aurora periods with \cora{other "planetary period" oscillations, \cord{or "PPO", which are, like the auroral periods, close to Saturn's rotation period.
One example is SKR, which we will compare with the auroral periods here.
Other examples are oscillations of the magnetic field observed by Cassini \citep{provan13}.}}
\cora{Saturn's magnetic field oscillates with two different periods \cord{which are close to Saturn's fluid "surface" rotation periods}.
The \cord{magnetic field} periods are related to the northern and southern polar regions and have different amplitudes.
These periods change with season.
Also, as measured by Cassini in 2009-2012  \citep{provan13}, they change abruptly  at $\sim$6- to 8-month intervals. 
Correlation of such periods and their changes with \corb{appearance and rotational periods of various} atmospheric phenomena (\eg, thunderstorms) suggests that the magnetosphere oscillations may be driven or influenced by deep neutral atmosphere of Saturn \citep{cowley13, fischer14}.}

\cora{\cord{Like in our study,} near-corotation was seen in Cassini observations of SKR, UV and IR aurora, and aurora-generated energetic neutral atoms \citep{lamy13}.
Near-corotation was also seen  in the cross-platform auroral observational campaign in spring 2013, which observed UV, visible, and IR aurora of both Saturn's poles \citep{melin14}.
\cite{badman12} shows that infrared aurora depends on local time, but also shows Ôplanetary periodÕ rotational modulation suggesting that aurora may be driven from inside SaturnÕs magnetosphere.}

\begin{figure}[htbp]
    \hspace{-0.5in}
    \rotatebox{90}{    
    \resizebox{!}{6.5in}{\includegraphics{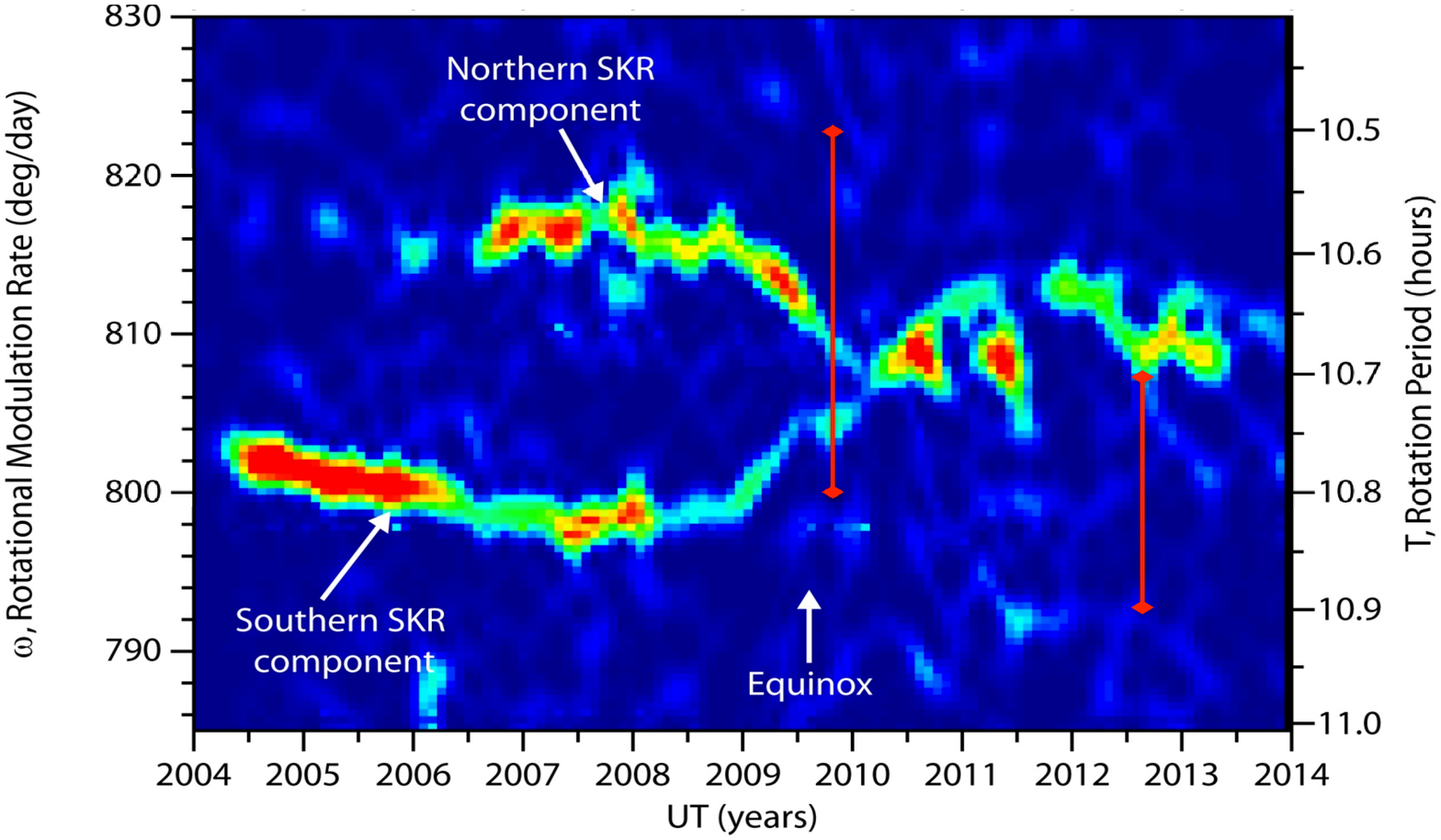}}
    }
   \vspace{-1.2in}
    \caption{
SKR periods varying with time during Cassini mission.
(D. Gurnett, private communication, an updated version of the plot by \cite{gurnett10})
The color scheme shows high SKR intensity in red, and low in blue.
Our aurora periods are plotted on top of SKR data as red error bars to show the uncertainty in the aurora period.
 }
    \label{fig: skr_period} 
\end{figure}

\cora{As an example, here we compare the auroral period with} the SKR period.
SKR period was used as a standard for Saturn's rotation rate since Voyager times.
\cora{However, from Ulysses \citep{galopeau00} and later observations it became clear} that the SKR period changes with time, and that there are different SKR periods in the North and South hemispheres.
Figure \ref{fig: skr_period} compares SKR periods with auroral periods that we have derived in this paper.
\renewcommand{\baselinestretch}{1.} 
\renewcommand{\baselinestretch}{2.}
\cora{Although consistent with both Northeren and Southern SKR,} the 2009 aurora period is closer to the period of the Northern SKR component.
This is not surprising because this observation shows the Northern aurora.
The 2012 aurora period is longer than any SKR period detected at the time, \cora{although it may be marginally consistent with \cord{the simultaneous SKR observation} given the large error bar on the aurora period}.
The 2012 aurora was observed at the South.
\cora{No direct comparison of visible aurora \cord{rotational phase and timing with simultaneous} SKR or magnetic field oscillations is performed here. 
Such comparison is an interesting subject for separate future research.
}

\cora{Physical discussion of the origin of near-corotation periods on Saturn has variously invoked spontaneous symmetry breaking in centrifugally driven outflow \citep{goldreich07}, longitudinal asymmetry in the ring current plasma \citep{khurana09, brandt10}, and perturbations driven by rotating wind systems in the polar thermosphere \citep{jia12,southwood14}, possibly influenced by major atmospheric storms \citep{cowley13,fischer14}. 
However, no consensus has yet emerged.
Continuing observations of aurora and comparison of visible observations with aurora-related observations by different instruments may help resolve this controversy.
}

\section{Spectrum and Vertical Structure.}
\label{sec:spectrum}

Movies taken with a sequence of different filters give information about the visible spectrum of Saturn{'}s auroras.
The multi-filter observations available to date are listed in Table \ref{tab: data}
\cora{\citep[see filter details in][]{porco04}.
}
The observation on  Nov. 27 (DOY 331) 2010 gives the best detection of aurora with nine filters.
We discuss this observation in detail.

\begin{figure}[htbp]
    \vspace{-2in}
    \resizebox{4in}{!}{\includegraphics{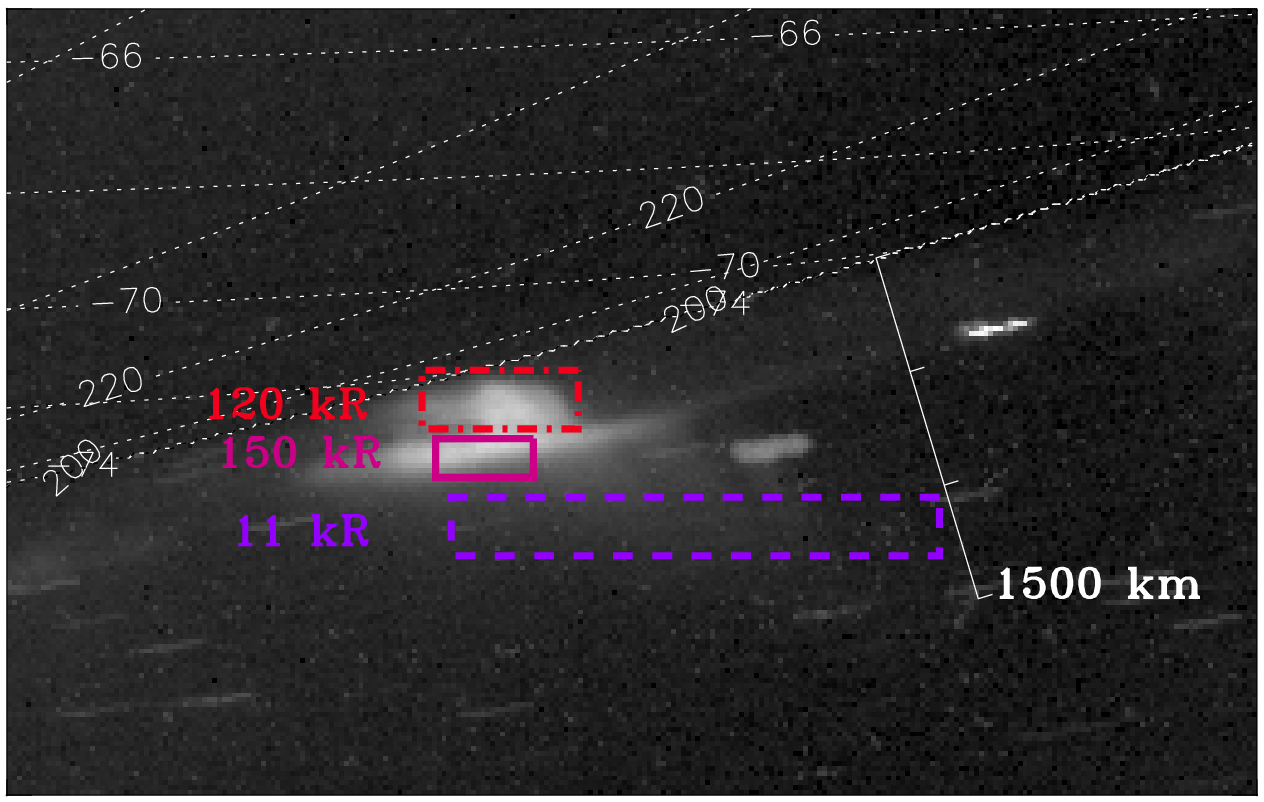}}
    \vspace{-0.05in}
    \newline
    \vspace{-0.15in}
    \hspace{-0.1in}
    \resizebox{4in}{!}{\includegraphics{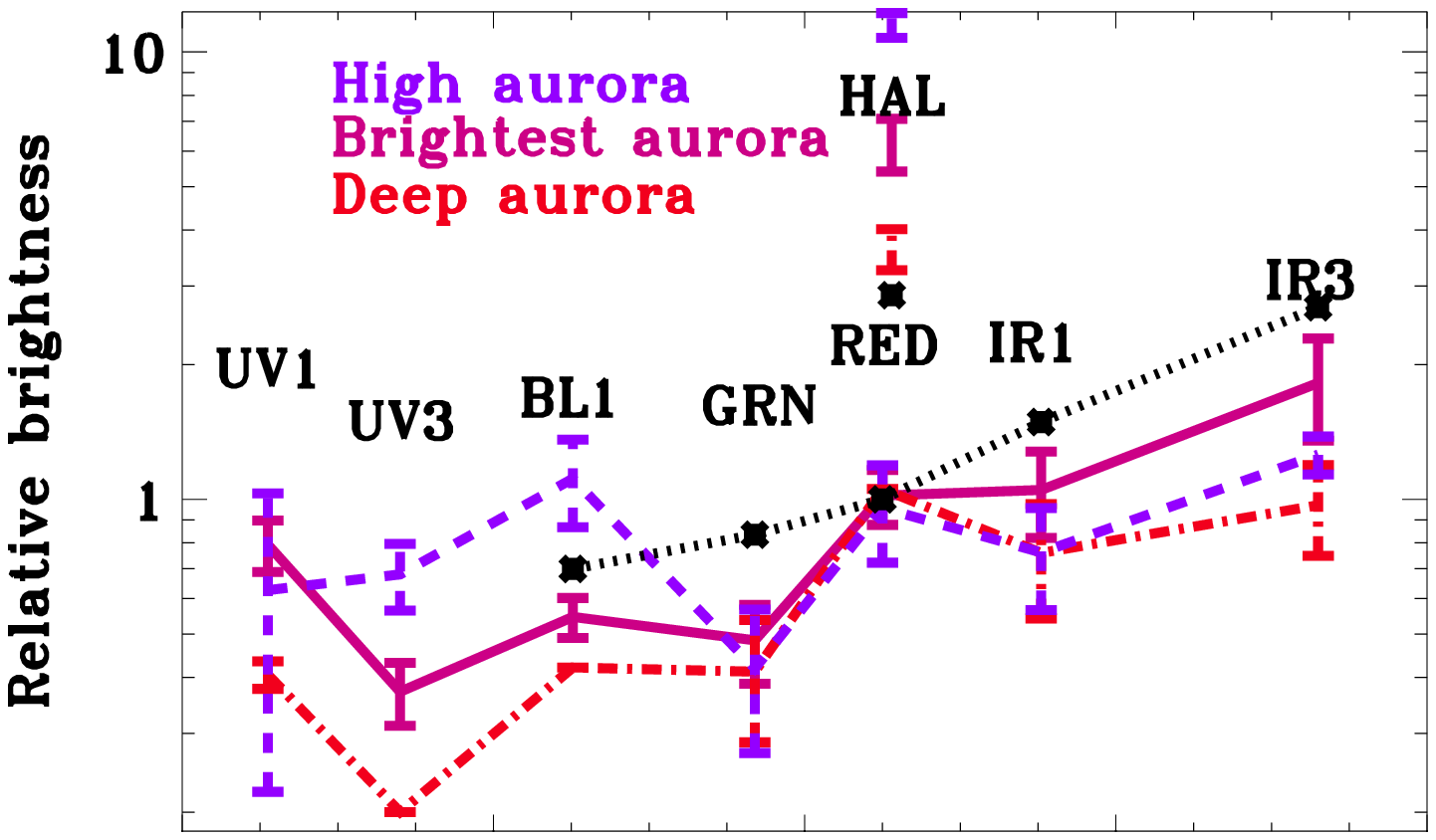}}    
    \newline
    \hspace{-0.1in}
    \resizebox{4in}{!}{\includegraphics{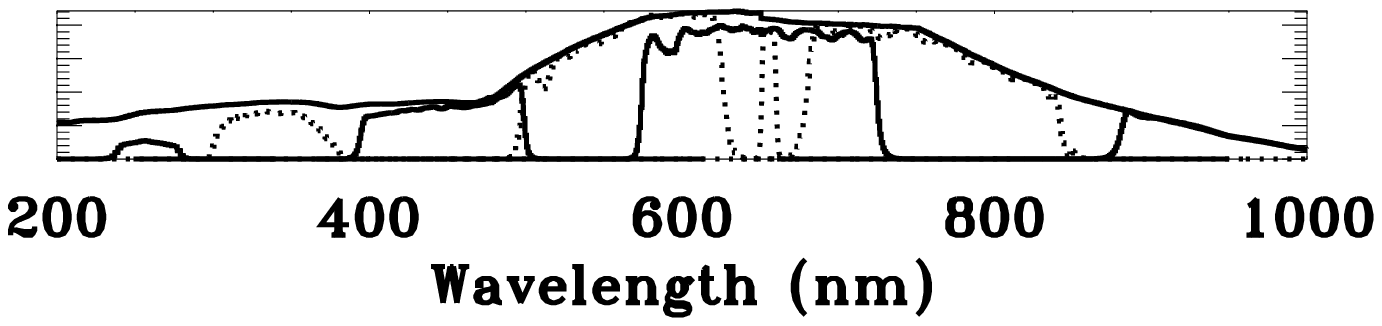}}    
    \newline
    \vspace{0.2in}
    \caption{
Upper panel: 
the image of aurora in 2010, DOY 331 sequence.
The longitudes and latitudes determined by default navigation are marked on the planet by the white dashed lines.
Star trails appear in the clear sky due to the spacecraft motion.
Areas where the spectra of the aurora were sampled with 9 filters are shown by the colored boxes.
The color of the boxes is the true color as derived from the measurements in RED, GRN and BL1 filters.
Middle panel: 
Colored curves show the auroral spectra measured with ISS filters (see filter shapes in the lower panel).
The curves' colors and plotting styles correspond to the boxes in the upper panel where the spectra were sampled.
Brightness in Rayleigh/nm in each filter is normalized by the brightness in the clear filter.
Note the logarithmic brightness scale.
The vertical error bars indicate the measurement's uncertainty.
Data points in broadband (50-200-nm-wide) filters \cora{are labeled with filter names and} are connected by lines. 
Narrow band H$_\alpha$ filter (\cora{labeled} HAL) brightness is shown by separate data points. 
The black dotted curve shows a laboratory simulated spectrum convolved with ISS filter shapes.
The simulated spectrum was provided by A. Aguilar, see also \cite{aguilar08}.
Lower panel:
\cora{Transmissivity} of the ISS filters \citep{porco04}.
The names of the filters are marked in the middle panel. 
Every other filter is shown by a dotted line to avoid confusion.
The $\sim$10-nm-wide HAL filter at 656nm is also shown by a dotted line.
}
    \label{fig:image_areas_spectra} 
\end{figure}

The upper panel of Figure \ref{fig:image_areas_spectra} shows a clear-filter image detecting a particularly bright aurora on this day.
\renewcommand{\baselinestretch}{1.} 
\renewcommand{\baselinestretch}{2.}
The dark nightside disk of Saturn near its south pole is on the upper left (South is down).
The aurora is seen above the limb on the star-streaked clear sky background (in the image the aurora is extending from the the limb downward).
The 9-filter spectrum in the middle panel is composed from the brightness measurements of multi-filter images sampled in the colored boxes of the upper panel. 
Filter transmissivity curves are shown in the lower panel.
The images of aurora in different filters were not taken simultaneously, but a few minutes apart.
\cord{This created a bias because }
the aurora changed substantially on that time scale.
To separate spectral and temporal variation, we used the time-interpolated brightness in the clear filter to \cord{derive normalized brightnesses} in other filters (see  details \cord{of interpolation} in Section \ref{sec: spectral_calibration}).

Among other filters, the aurora was detected in red, green, and blue.
This allowed us to determine the true color of the aurora (see details in Section \ref{sec: spectral_calibration}).
The color of the sampling boxes in the upper panel indicates the true color.
Accordingly, the auroral curtains change from pink at a few hundred km above the horizon \corb{(dot-dashed box and spectrum in Fig. \ref{fig:image_areas_spectra})} to purple at 1000-1500 km above the horizon \corb{(dashed box and spectrum)}.
The auroral curtains on Earth are green at the bottom and red at the top 
due to excited nitrogen and oxygen atoms and molecules.
Saturn's auroras are dominated by excited forms of hydrogen - the main constituent of Saturn atmosphere. 
Colors can differ due to changes of atmospheric density, the chemical state of the element, and the energy of impacting particles.
\cora{For example, the brightest green aurora on Earth is due to oxygen in atomic form.
The rapid decrease of concentration of atomic oxygen below about 100 km is responsible for the abrupt-looking bottom parts of the Earth's auroral curtains.
\cord{The bottom of the auroral curtain in Fig. \ref{fig:image_areas_spectra} is red}.
Like on Earth, some reddening of the aurora near the horizon may be due to atmospheric scattering of the light on the way to the observer.}
\cord{This paper reports the first detection of visible-light auroral spectrum on a planet other than Earth. 
Visible aurora detected on Jupiter by Galileo spacecraft was only observed in one broadband channel \citep{vasavada99}}.

The 9-filter Saturn's aurora spectrum can tell us about Saturn's upper atmosphere and the electrons impacting on it.
Laboratory simulations or modeling of the spectrum are needed to \cord{learn about atmospheric conditions and impacting particles from the} observed vertical distribution of the spectra.
The results of one such simulation  \citep{aguilar08} are shown in the middle panel of Fig. \ref{fig:image_areas_spectra}.
\corb{We convolved the high-spectral-resolution laboratory spectrum with Cassini filter shapes (bottom panel).}
The resulting black \cord{dotted} curve shows the electron impact-induced fluorescence spectrum of H$_2$ for electron energy 20 
eV.
\cora{This energy \cord{may be relevant to aurora because it} is typical for secondary electrons, which contribute substantially to UV aurora on Jupter \citep{ajello05} and airglow on Saturn \citep{gustin10}.}

The bright H$_\alpha$ (656nm) line is the strongest emission line of the laboratory simulation.
\cora{This can be seen in the middle panel of  Fig. \ref{fig:image_areas_spectra}, where the H$_\alpha$ "HAL" data point (separate black {$\times$} symbol) is substantially brighter than the average for these wavelengths broadband "RED" data point (black {$\times$} symbol laying on the black dotted line).}
The observed aurora has an even stronger H$_\alpha$ line than the simulated one.
\cora{This can be seen in the middle panel of  Fig. \ref{fig:image_areas_spectra}, where} the colored \cora{separate "HAL" data points from the observation are brighter than the black separate data point from the simulation \corb{(the black $\times$ symbol off the black line)};} note the log scale for brightness.
The laboratory simulation shows the same general "red" \cora{slope in the spectrum} as our observations, though there are differences.
For example, the low brightness of the observed aurora in the green filter, which produces the pink-purple aurora color, is not seen in the lab spectrum.
Broader ranges of modeled atmospheric properties and impacting particles \cord{need to be modeled to}
\cord{derive atmospheric structure an particle properties from }the observed spectrum. 

Comparing the aurora in Fig. \ref{fig:image_areas_spectra} \cora{with} the scale bar in the upper panel, one can see that the auroras extend from a few hundred to $\sim$1500 km above the horizon. 
The horizon is defined by Saturn's atmosphere blocking the star trails.
This is close (within 20-50 km) to the 1-bar limb defined by the default Cassini navigation, which is subject to a $\sim$50-km navigation error. 
The UV aurora was observed by HST at  900-1300 km above the 1-bar level. 
Its UV spectrum suggests altitudes above 610 km  \citep{gerard09}.
This \cord{may mean}  higher elevations of the UV aurora.
Indeed, our spectrum shows a larger contribution from UV wavelengths in the high-altitude aurora than in the low-altitude aurora. 
\cord{However UV observations from HST have lower spatial resolution.}
Accordingly, the elevation difference may be due to \cord{sampling uncertainty or due to higher elevation of UV aurora than the visible auroras}.

\cord{On Jupiter visible aurora appears at altitude 245$\pm$30 km above the limb at its brightest part, and extends up by 120$\pm$40 km \citep{vasavada99}.
}
Terrestrial auroral curtains usually extend from $\sim$100 km up to 200-300 km altitude.
The larger altitudes of Saturn's aurora are probably due to Saturn's more extended atmosphere, where the density decreases with height about 10 times slower than in the Earth's atmosphere \cora{due to the lower molecular mass} \cord{and lower gravity than Jupiter}.

\section{Conclusions}

\cora{Visible aurora on Saturn was first detected on the night side of Saturn by Cassini camera, and continues to be observed.
The auroral movies and maps from the first detection in 2006 till 2013 are reported in this paper.}

\cora{These observations reveal auroral colors changing with height from pink to purple.
The visible auroral spectrum was sampled by seven 50-200-nm-wide filters spread across 250-1000 nm, a broadband filter covering the whole range of wavelengths, and the 10-nm-wide H$_\alpha$ filter.
Aurora shows prominent H$_\alpha$ line emission.}
\cord{This is the first detection of visible aurora spectrum on a planet other than Earth.}

\cora{Visible aurora observations have unprecedented spatial (down to 11 km/pixel) and temporal (down to 1 minute step in the movies) resolution.
This allowed us to  detect near-corotation of aurora with Saturn.
With two long observations we derived rotation periods of auroral structure in 2009 and 2012.
These periods are within the range of other planetary period oscillations detected on Saturn.
Also we detected sudden few-minute-long brightenings of aurora repeating every $\sim$ 1 hour, and changes in small aurora features' corotation speed of the timescales of minutes in some locations on auroral oval. 
}

\cora{The location of the main auroral oval is 70-75{\deg}  North or South latitude, similar to UV and IR auroras.
As observed near midnight, the oval changes latitude as Saturn rotates, forming a bright spiral segment in System III \core{coordinate} map projection, which \cord{persists} on consecutive \cord{Saturn} rotations.
The bright segment's center of rotation is displaced from the pole in local time in anti-solar or dawn direction, as judged by by the dusk-midnight sector observed.}

\appendix
\beginappendix
\section{Appendix}
\subsection{List of Aurora Detections}
\label{sec: aurora_detections}

The aurora was detected by the Cassini camera using two approaches.
One was to take images with long exposures to build up signal to noise in one image.
Another approach, which proved itself more effective as it detected more aurora, was to make multiple short exposures, which were later combined into movies.
The "movie-style" observations can be done at the expense of spatial resolution by onboard pixel binning.
Movies allowed us to detect auroral motion and to get spectral information from different filters closer in time than with long exposures.

Table \ref{tab: data} summarizes all ISS observations that have detected Saturn{'}s aurora to date.
\renewcommand{\baselinestretch}{1.} 
\begin{table}[htbp]
\caption{Cassini camera observations which detected aurora.
The start time of the observation is given as follows: (year)-(day of the year)T(hour):(minute):(second).
Duration of the image sequence is given as (hour):(minute) from the start of the first image to the start of the last image.
In some observations, the sequence of  $N_{filt}$ images in different filters was 
repeated N$_{fr}$ times. 
For single-filter movies N$_{fr}$ is the total number of images.
"Latitudes" lists the range of observations as the maximum and minimum latitudes covered by at least one image in the data set.
The "km/pix" column gives the image scale in km at the end of the observation, not accounting for slant-viewing foreshortening.
The asterisks (*) mark noisy detections.
\cora{The crosses in "UV" and "IR" columns indicate simultaneous with visible observations Cassini ultraviolet and/or infrared aurora observations, respectively.}} 
\vskip4mm
\centering
\begin{tiny}
\begin{tabular}{|c|c|c|c|c|c|c|c|}
\hline
Start time & Duration &Filters [exposures in sec.] & $N_{fr}\times N_{filt}$  &  Latitudes & $km\over pix$ & UV& IR\\
\hline 
2006-197T01:32:14&10:43 & CLR[150,68],& 5$\times$9 & 67 to 89 & 15 & & \\
& & UV1, UV3, IR2, HAL [1200,560] &  & & 15 & & \\
non-detections 	& & MT2,MT3,CB2,CB3[1200,560]&  &  & 15 & & \\
\hline
2007-003T06:54:25&10:15 & CLR[68] & 9$\times$1 & 61 to 77 & 16 & & x\\
2007-003T07:00:44&10:15 & CLR[68] & 9$\times$1 & 68 to 80 & 16 & & \\
\hline
2009-023T08:03:43*&2:04 &  CLR[150,26,120,680] & 12$\times$1 & 71 to 76 & 6 &x & x\\
\hline
 2009-278T16:50:09&9:03 & CLR[180] & 164$\times$1 & 54 to 80 & 32 &x & \\
\hline
2009-279T15:18:09&8:30 & CLR[180] & 148$\times$1 & 54 to 80 & 31 &x & \\
\hline
2009-280T20:49:08&4:57 & CLR[180] & 87$\times$1 & 55 to 79 & 29 &x & \\
\hline
2009-281T21:49:09&4:00 &  CLR[120] CLR[180] & 73$\times$1 & 55 to 79 & 27 &x & \\
 \hline
 2010-177T04:18:44&12:37 & CLR[38] & 628$\times$1 & -79 to -59 & 52 &x & \\
 \hline
 2010-179T04:37:26*&4:30 & CLR[150] & 10$\times$9 & -79 to -59 & 26 &x & \\
 				&	 & RED[100] ,HAL[180]& 	 &  & 52 & & \\
non-detections 	&	 & UV3,UV1,BL1,IR3[180,]&  &  & 52 & & \\
non-detections 	&	 & IR1[100],GRN[120]&  &  & 52 & & \\
 \hline
2010-180T13:13:26&10:30 & CLR[150] & 22$\times$9 & -78 to -60 & 25 & x& \\
	&	& RED,IR1[100] ; BL1,HAL[180] & 	 & 	& 50 & & \\
non-detections 	&	 & UV3,UV1,IR3[180],GRN[120]&  &  & 50 & & \\
 \hline
2010-331T01:13:19&4:30 & CLR[150] & 10 $\times$9 & -76 to -53 & 20 & x& x\\ 
& & BL1,HAL,IR3,UV1,UV3[180] & & -76 to -54 & 40 & & \\
& & RED,IR1[100],GRN[120] & & -76 to -54 & 40 & & \\
 \hline
 2011-028T12:56:06&4:30 & CLR[150] & 10$\times$9 & -86 to -57 & 19 & x & x\\
& & RED,IR1[100],HAL[180] &  &  & 38 & & \\ 
non-detections& & UV3,UV1,IR3,BL1[180] &  &  & 38 & & \\
\hline
2012-067T19:24:36*&4:55 & CLR[0.68] & 737$\times$1 & -75 to -57 & 29 &x & \\
\hline
2012-136T15:45:52*&6:01 & CLR[0.68] & 902$\times$1 & -77 to -54 & 47 &x &x \\
\hline 
2012-137T11:16:12*&5:06 & CLR[0.68] & 765$\times$1 & -77 to -55 & 42 &x &x \\
\hline 
2012-195T07:35:03&16:33 & CLR[32] & 266 & -88 to -61 & 66 & x& x\\
\hline
 2012-197T22:49:02&11:27 & CLR[32] & 180 & -85 to -60 & 59 & x& x\\
\hline
2012-199T12:33:24&11:15 & CLR[38] & 277$\times$1 & -83 to -60 & 53 & x&x \\
\hline
2012-244T06:28:33*&7:24 & CLR[18] & 120$\times$1 & -78 to -64 & 35 &x &x \\
\hline
2012-329T02:59:56&4:29 & CLR[18] & 289$\times$1 & -80 to -60 & 31 & x&x \\
\hline 
2012-342T21:45:39&6:40 & CLR[150] & 98$\times$1 & -76 to -66 & 11 & x&x \\
\hline 
2013-003T16:48:39&2:20 & CLR[150] & 43$\times$1 & -76 to -66 & 11 & x& x\\
\hline 
2013-078T20:57:03*&5:14 & CLR[38] & 263$\times$1 & -75 to -67 & 22 &x &x \\
\hline 
2013-110T15:43:24*&3:02 & CLR[38] & 100$\times$1 & -75 to -68 & 18 & x& x\\
\hline 
2013-111T08:40:02&5:49 & CLR[32] & 109$\times$1 & -73 to -69 & 11 &x & x\\
\hline     
      \end{tabular}
      \end{tiny}
    \label{tab: data}
\end{table}
\cora{
Aurora-targeted observations which did not result in detection are not listed.
Some of the observations were taken with several filters. 
The filters in which aurora was detected (or not detected) are listed in Table \ref{tab: data} by the filter names (see filter details in \cite{porco04}).
These filters' spectral shapes are also shown in Fig. \ref{fig:image_areas_spectra}. 
}

\cora{For the reference of  future cross-instrument studies, we list simultaneous or nearly simultaneous (within few hours) auroral observations by two other Cassini instruments:
 UVIS (ultraviolet wavelengths, labeled UV in Table \ref{tab: data}) and VIMS (infrared wavelengths, labeled IR). 
}

\subsection{Spectral calibration}
\label{sec: spectral_calibration}

\begin{figure}[htbp]
    \resizebox{5in}{!}{\includegraphics{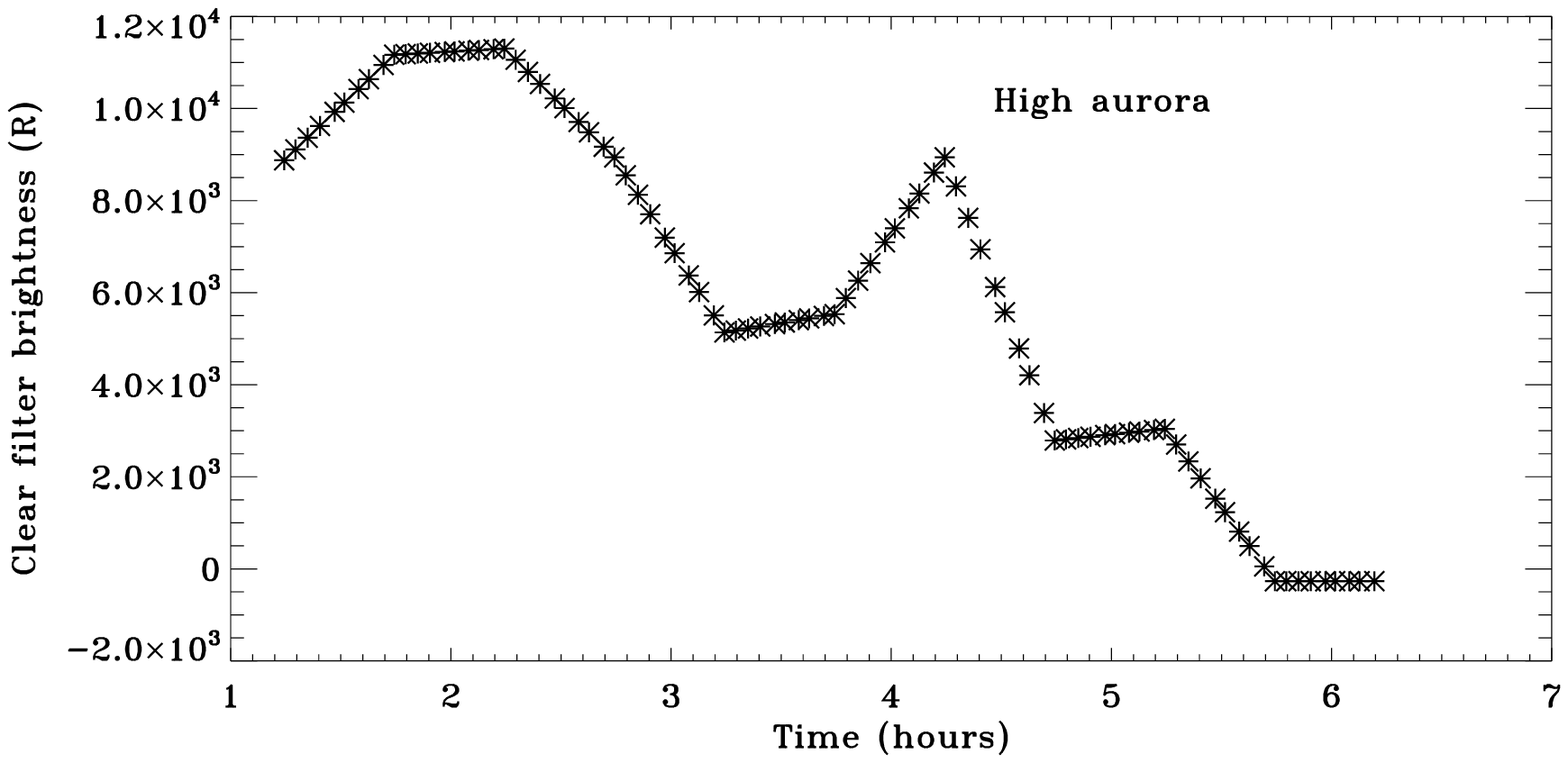}}
    \vspace{0.2in}
    \caption{
Clear-filter brightness changing with time on  Nov. 27 2010 (2010-331 observation in Table \ref{tab: data}).
The kinks in the curve show the data points from the clear filter images, which are obtained  about every half--hour.
The brightness between the points is linearly interpolated to be used to normalize the brightness of the other filters at the times these observations were taken.
The asterisks indicate times when observations were taken in different filters.
The clear filter is assumed to have an effective width of 600 nm. 
} 
    \label{fig:brightness_versus_time} 
\end{figure}

To obtain the spectrum of the auroras, we divided the auroral brightness (in units of R/nm) in all filters by the brightness in the clear filter.
Because observations are not simultaneous, we interpolate changes in the clear-filter auroral brightness with time at each auroral location (color box in Fig. \ref{fig:image_areas_spectra}). 
Figure \ref{fig:brightness_versus_time} shows the variable brightness of the high-elevation part of the aurora in the purple box in Fig. \ref{fig:image_areas_spectra}.
\renewcommand{\baselinestretch}{1.} 
\renewcommand{\baselinestretch}{2.} 
The brightness in filters other than clear is then divided by the interpolated value of the clear filter taken at the time of observation.
This produces a set of 10 normalized brightnesses (1 per time step) for each filter (the observation consisted of 10 repeated observations with 9 filters each).
At some time steps in some filters the aurora was too faint to detect.
Using only the time steps when the aurora was detectable in each non-clear  filter, we averaged these normalized brightnesses to provide a data point for Fig \ref{fig:image_areas_spectra}. 

The true colors were determined by calibrating images into units of reflectivity $(I/F)$ instead of Rayleighs (R).
The $I/F$ units normalize the detected brightness $I$ by the brightness a white Lambertian surface at Saturn's distance from the Sun $F$.
Combination of red, green, and blue brightnesses in $I/F$ units was assumed to represent the true color observed by the human eye.

\ack
We thank Peter Goldreich for fruitful discussions. 
We thank Valeria Verkhovyh for participating in the aurora data analysis.
We thank Joseph M. Ajello and Alex Aguilar for providing lab-simulated auroral spectra.
We thank Don Gurnett for providing the SKR data.
This research was supported by the NASA Cassini Project.

\label{lastpage}


\bibliography{my}

\bibliographystyle{icarus}


\clearpage	

\clearpage

\beginsupplement
\section*{Online Supplementary Data Captions}

The movies can be found at  http://web.gps.caltech.edu/ $\sim$ulyana/www\_papers/aurora/SOM/

\renewcommand{\figurename}{Supplementary Movie}

\begin{figure}[htbp]
    \caption{\corb{Movie of aurora observed on July 8 (DOY 199) 2012 processed similarly to Fig. \ref{fig: unprojected}.
There are 277 frames in the movie (see more details in Table \ref{tab: data} of Section \ref{sec: aurora_detections}).
An average of all image frames was subtracted from each of the frames to remove background.
West longitude and planetocentic latitude grid overlays the images. 
The date and time are labeled in the lower left of each movie frame. 
Each image was taken in CL1+CL2 filter combination.
\core{One degree of latitude is about 1000 km.}}
} 
    \label{fig: movie199_12} 
\end{figure}

\begin{figure}[htbp]
    \caption{Movie of aurora observed on June 26 (DOY 177) 2010 processed similarly to Fig. \ref{fig: unprojected} and Supplementary Movie \ref{fig: movie199_12}. 
There are 628 frames in the movie.
An average of frames 300 through 400 was subtracted from each of the frames to remove background.
Each image was taken in CL1+CL2 filter combination.
} 
    \label{fig: movie177_10} 
\end{figure}

\begin{figure}[htbp]
    \caption{Movie of aurora observed on  November 24 (DOY 329) 2012 processed similarly to Fig. \ref{fig: unprojected} and Supplementary Movie \ref{fig: movie199_12}.   Each image was taken in CL1+CL2 filter combination.} 
    \label{fig: movie329_12} 
\end{figure}

\begin{figure}[htbp]
    \caption{Movie of aurora observed on on  October 6 (DOY 280) 2009 processed similarly to Fig. \ref{fig: unprojected} and Supplementary Movie \ref{fig: movie199_12}.   Each image was taken in CL1+CL2 filter combination.} 
    \label{fig: movie280_09} 
\end{figure}

\begin{figure}[htbp]
    \caption{Movies obtained on October 4-7 (DOY 278-281) 2009 (see more details in Table \ref{tab: data} of Section \ref{sec: aurora_detections}).
The images are processed the same way as in Supplementary Movies \ref{fig: movie199_12} -  \ref{fig: movie280_09}.
The timing of the movies can be seen in Fig. \ref{fig: doy278_09_lat_plot}.
Parts of the four movies of the same area taken on these dates are aligned such that their starts are separated by several Saturn whole rotations assuming a rotation period of 10.65 hours.
The movie demonstrates auroral structures persisting at the same location several Saturn rotations apart.} 
    \label{fig: parallel_movie281_09} 
\end{figure}

\begin{figure}[htbp]
    \caption{Movies obtained on July 4-8 (DOY 195-199) 2012 (see more details in Table \ref{tab: data} of Section \ref{sec: aurora_detections}).
The movies are aligned the same way as in  Supplementary Movie \ref{fig: parallel_movie281_09} but assuming a rotation period of 10.8 hours.
The timing of the movies can be seen in Fig. \ref{fig: doy195_12_lat_plot}.
On DOY 195 at time $\sim$ 20:40 the bright moon Mimas can be seen moving across the sky.
} 
    \label{fig: parallel_movie199_12} 
\end{figure}

\end{document}